\documentclass[12pt]{article}
\usepackage{amsmath, amssymb,booktabs}
\usepackage{hyperref}
\usepackage{geometry}
\usepackage{tikz}
\usetikzlibrary{positioning}
\title{Social Laser Theory as a Natural Extension of Quantum-Like Modeling}
\author{Andrei Khrennikov \\Center for Mathematical Modeling in Physics and Cognitive Sciences\\Linnaeus University, V\"axj\"o, SE-351 95, Sweden}

\begin{document}
\maketitle
\abstract{This paper presents a comprehensive review of the \emph{Social Laser Theory} (SLT) as a natural extension of the broader framework of \emph{Quantum-Like Modeling} (QLM). While QLM applies the mathematical formalism of quantum theory—such as Hilbert space representations, interference, and non-commutative observables—to model context-dependent and non-classical phenomena in cognition, decision-making, and social behavior, SLT advances this approach by integrating concepts from quantum field theory. The theory conceptualizes social systems as ensembles of ``social atoms’’ capable of absorbing and emitting quantized units of social energy. Under conditions analogous to population inversion in physical lasers, external informational stimuli (e.g., media signals or mobilizing rhetoric) can trigger coherence across the population, resulting in large-scale, synchronized collective actions such as protests or ideological shifts. SLT thus provides a formal framework for understanding the amplification and coherence of social energy leading to abrupt phase-like transitions in collective behavior. Beyond its metaphorical appeal, the theory proposes measurable quantities and predictive parameters that may support empirical diagnostics of sociopolitical dynamics. By bridging micro-level psychological processes with macro-level sociological phenomena, SLT extends QLM into the domain of complex social systems, offering a mathematically grounded paradigm for interpreting rapid transformations in contemporary societies.}

{\bf keywords:} quantum-like modeling, social laser, quantum probability and information, quantum-like cognition and decision making, sociaopolitical dynamics, social  phase transition, discursive frame, cognitive marker 

\section{Introduction}

This paper is a review on the {\it Social Laser Theory} (SLT) presented as a natural counterpart of {\it quantum-like modeling} (QLM). 

In recent years, researchers across disciplines have become increasingly aware that many aspects of biology (genetics, epigenetics, evolution theory),  human cognition, collective behavior,  social and political dynamics, economics and finance, game theory resist adequate description within the framework of classical probability theory and logic (which were formalized by Kolmogorov \cite{K} and Boole \cite{Boole1,Boole2} respectively). Phenomena such as interference and context-dependence in decision making, violations of additivity in probability judgments, or sudden large-scale social shifts - social tsunamis - highlight the limitations of traditional analytical tools. These challenges have stimulated the rise of an interdisciplinary approach often referred to as QLM (see 
pioneer works \cite{Aerts1}-\cite{Aerts3}, \cite{KHC1}-\cite{INT,QL0}, see also some selection of references in QLM 
\cite{UB_KHR}-\cite{PLOS}, \cite{Bruza}-\cite{Busemeyer5}, \cite{Pothos1,Pothos2,Asano,QIB,QL3,Behti,BioBas}, \cite{Khrennikov}-\cite{Polina1}, \cite{Bagarello1}-\cite{Bagarello4}, \cite{QLH,OK20,OK21,OK23,Fuyama}, \cite{HavenB}-\cite{Polina2}, \cite{Gunji1}-\cite{Gunji4}). Rather than suggesting that human thought or society is governed by quantum mechanics in a literal physical sense, QLM adopts the {\it formal structures of quantum theory}---notably complex Hilbert space representation, interference (violation of the formula of total probability), superposition, and complementarity/incompatibility of observables  
(non-commutativity of operators representing observables)---as flexible mathematical resources for capturing non-classical dependencies in information processing and collective action.  QLM is one of the co-products of the quantum information revolution and nowadays one can speak about {\it the quantum-like revolution} \cite{Zeit}). As the result of the quantum information revolution, the quantum community started to be more polite to applications of the formalism and methodology of quantum theory beyond quantum physics.

This is the good place to note that QLM in cognition should be sharply distinguished from the ambiguous project towards reduction of mental phenomena to quantum physical processes in the brain, see, e.g., \cite{H,P, V1,V2,V3, Igamberdiev2,Igamberdiev2a,Igamberdiev1b,Melkikh3}. 
This project can be called  the {\it quantum consciousness}  project: at least in the works of Hameroff \cite{H} and Penrose \cite{P} and other authors developing their model this term was actively used and indirectly in the works of Vitiello \cite{V1,V2} as ``quantum brain model'' (see also  \cite{V3}).   The quantum consciousness project is directly related  to quantum biophysics \cite{QBIOP,Igamberdiev1,Melkikh1,Melkikh2}. Quantum biophysics  deals directly with quantum physical processes in biological systems, QLM focuses on macroscopic biological, social, economic and financial and, more recently, AI systems. QLM describes information processing in such systems  using quantum information and probability principles, though it is not directly rooted in quantum physics.\footnote{ 
It seems to be natural to couple the terminology ``quantum'' to models based on genuine quantum physics and ``quantum-like'' to applications of quantum probability, information, and quantum measurement theory  beyond physics, e.g., to mental phenomena.   
But the terminological landscape is even more complicated, a  few  leading cognitive scientists, as Aerts, Broekaert, Bruza, Busemeyer, Phothos, Smets have been using the term ``Quantum Cognition'', instead of ``Quantum-like Cognition'', although they didn't try connect mental phenomena with quantum physical processes in the brain.}   
 
Within this broader, QLM, movement, an especially ambitious framework is SLT, first articulated by the author \cite{laser1,laser2} and subsequently elaborated with several coauthors \cite{laser3,laser4,laser5,laser6,laser7,laser8,laser9,laser10,laser11,laser12}. SLT introduces concepts from {\it quantum field theory}, extending QLM beyond finite-dimensional models toward systems with infinitely many degrees of freedom (cf. \cite{V1,V2,V3}). The central metaphor is drawn from physics: just as a physical laser relies on the amplification of coherent photon flows through stimulated emission, a ``social laser'' describes how informational and emotional excitations can accumulate in a population until collective, coherent action is released. In this picture, individuals---conceived as {\it ``social atoms''}---occupy quantized energy-like states, absorbing and emitting informational ``quanta.'' Under conditions of {\it population inversion} and with the right external stimulus (for instance, repeated media signals or mobilizing rhetoric), the system may undergo a phase-like transition from dispersed attitudes to synchronized mass behavior.  

Nowadays development of SLT is especially important, since mass social protests disturb even well organized democratic societies, in 
Western Europe and USA (as the series of mass protests against policies of Macron and Trump). SLT can give hints to prevention of social explosions. In future by developing experimental frameworks for  determination of the  basic parameters, e.g., approaching the stage of population inversion, SLT can provide diagnostics of the sociopolitical situation.     

In connection to SLT we definitely should mention  contribution of Haken  \cite{Haken1983}. Although he did not introduce the term 
\emph{social laser} or formulate a detailed model based on laser physics for social systems,  he employed \emph{laser-like metaphors} to explain general mechanisms of self-organization and coherence. His authority in laser physics and similarity in thinking about 
laser-like social phenomena are very supportive for our social laser project (see section \ref{Haken}).  

The notion ``social energy'' and ``social atom'' are basic in SLT. As well as energy in quantum physics, social energy is treated operationally as a quantity determinable via special measurement procedures. Thus, this notion is only indirectly coupled to such notions as {\it mental and psychic energy} considered in the works of James \cite{James}, Freud \cite{Freud_sex1,Freud_sex2,Freud_pleasure1,Freud_pleasure2}, and Jung \cite{Jung}.  They tried to go deeper - beyond the operational approach. But, honestly saying their presentation of  mental and psychic energy is fazzy and only metaphorically related to physical energy (section \ref{energy}).

The choice of the term ``laser'' is not merely decorative. Early formulations even considered the broader label of an {\it information laser} \cite{laser1}. Yet ``social laser'' emphasizes the specifically societal dimension, while retaining the structural analogy with physical lasers: pumping fields, excitation, coherence, and emission. In this sense, one may interpret this mechanism as {\it  stimulated emission of social energy}, although the more concise ``social laser'' has been retained in the literature for clarity and resonance with its physical counterpart.  

Applications of this framework are diverse. At its core, SLT offers a way to conceptualize sudden, large-scale mobilizations---protest waves, political revolutions, or ideological cascades---that emerge not gradually but abruptly, once coherence conditions are met. The same formalism, however, can be extended to other collective domains, such as epidemic dynamics (``epidemic lasers'' \cite{covid}), financial crises (``financial lasers''), or coordinated responses to global challenges. What unites these examples is the underlying mechanism: the reorganization of distributed informational and emotional energy into coherent, collective patterns of action.  

The theoretical contribution of SLT lies in bridging {\it micro-level psychology} with {\it macro-level sociology} through a single formal lens. Whereas traditional models---such as rational-choice frameworks, threshold models, or contagion dynamics---tend to emphasize either gradual accumulation or local interactions, SLT introduces the possibility of rapid, system-wide synchronization grounded in non-classical probabilistic dynamics. In this sense, it extends the scope of QLM beyond individual cognitive puzzles (e.g., order effects in psychology) toward systemic accounts of social transformation.  

In what follows, we situate SLT within the landscape of quantum-like approaches, highlighting both its conceptual innovations and its differences from earlier models. We then examine its formal underpinnings, its analogies to quantum optics, and its explanatory reach relative to established theories of mobilization. Alongside, we discuss ongoing debates about terminology, operationalization, and empirical testability. Our aim is to present SLT not as a speculative metaphor, but as a structured and mathematically motivated framework that can enrich our understanding of contemporary social dynamics in an era of accelerated communication and high emotional resonance.  

The mathematical formalism of quantum theory will be briefly described in section \ref{QFDM}. 

\tableofcontents
\newpage

\section{Quantum-like Modeling}

\subsection{From Classical Probabilities to Quantum Structures}
The development of QLMs stems from fundamental limitations in classical models of uncertainty, information processing, and decision-making. Classical probability theory rooted in the Kolmogorovian measure-theoretical framework \cite{K} is built on assumptions of additivity, distributivity, and static sample spaces. These assumptions, while powerful in modeling idealized systems of dice and coins, often fail to capture the contextuality, indeterminacy, and interference effects observed in real-world cognition and behavior, individual and collective.

In numerous empirical studies across psychology and decision sciences, human judgments systematically violate the basic principles of 
classical Kolmogorovian probability theory.  Notable examples include the disjunction effect, order effects, conjunction fallacy, and question framing biases. These anomalies suggest that the mental processes underlying decision-making are not easily representable as a traversal through a fixed decision tree or a well-defined Kolmogorov probability space. 

QLM proposes a fundamental shift: rather than forcing human thought into a classical probabilistic framework, it employs the formalism of quantum mechanics, particularly the structure of Hilbert spaces, non-commutative observables, and amplitude-based interference, as a descriptive and predictive tool. Importantly, this approach does not posit that the brain is a quantum physical system, but rather that quantum mathematics and methodology provides a better syntax for modeling certain types of information flow, ambiguity resolution, and probabilistic reasoning in complex agents and systems.

\subsection{Core Features of the Quantum-like Formalism}
At the heart of QL modeling lies the representation of cognitive (mental, belief, or psychic) states as normalized vectors (determined up to a phase factor)  belonging to  a complex Hilbert space ${\cal H},$ and observables (e.g., decisions, tasks, measurements, or perceptions) as Hermitian operators acting on these states. More generally, the states are represented by density operators and observables by positive operator valued measures (POVMs). 

The transition from one cognitive state to another is captured by the projection postulate (L\"uders postulate), analogous to measurement in quantum mechanics. Probabilities arise from squared complex amplitudes, giving rise to quantum interference patterns.

Key features of this formalism include:
\begin{itemize}
\item Superposition: Cognitive states can exist in a linear combination of potential outcomes, reflecting uncertainty and ambiguity before resolution.
\item Contextuality: Measurements (questions, decisions) do not merely reveal a pre-existing state, but actively shape the outcome. The result depends on the total context, not just on hidden variables.
\item Non-commutativity: The order of questions or observations affects the results - mirroring empirical findings in survey responses and decision-making.
\item Entanglement: Composite systems (e.g., social groups, teams, societies) can exhibit joint states that are not reducible to individual states, reflecting interdependence and correlation across agents.
\end{itemize}

These features make QL models particularly suitable for systems characterized by cognitive overload, uncertainty, and non-classical dependencies, including social networks, political behavior, market dynamics, and collective action.

\subsection{Quantum formalism for decision making}
\label{QFDM}

Now we  describe - just i a few words - the QL framework for  decision-making. 
For simplicity, we consider the finite-dimensional case as is done in the majority of papers on QLM. 
Let ${\cal H}$  be a complex Hilbert space with the scalar product $\langle \cdot|\cdot\rangle$ and the corresponding norm $||\psi||^2= \langle \psi|\psi\rangle.$  Denote the set of density operators in ${\cal H}$ by the symbol ${\cal D}= {\cal D}({\cal H}).$  These are positive Hermitian trace one operators.   We remark that by commonly used terminology density operators represent mixed quantum states, statistical mixtures of pure states. 

The QL  states of a cognitive system, decision maker, are represented by normalized vectors of a complex Hilbert space 
${\cal H}$ or more generally by density operators. So, ${\cal D}$ is the space of QL states. QL observables -
questions, tasks, traits - are represented by Hermitian linear operators, $A:  {\cal H} \to {\cal H}.$ The spectrum of $A$ quantifies possible outcomes of observations, cognitive measurements.  For finite dimensional space  ${\cal H},$  the spectrum of $A$ is discrete and consist of eigenvalues, $x_1,...,x_n$  labeling the outcomes of observations, e.g., the possible answers to question $A.$
One can operate with matrices giving representation of operators in orthonormal bases in ${\cal H}.$ 
By the Born's rule the average of observable given by a Hermitian operator $A$ w.r.t. a state given by a density operator $\rho$  is given by the formula
\begin{equation}
\label{DM}   
\langle A\rangle_\rho = \rm{Tr} \rho A.
\end{equation}
One of the main distinguishing features of this rule is its linearity w.r.t. the state and observable (similarly to 
classical Kolmogorov probability theory \cite{K} in that averages are given by integrals w.r.t. probability measures). 

This rule is a simple probabilistic consequence of the Born's rule for probability. We note that any Hermitian operator 
$A$ can be represented in the form:
\begin{equation}
\label{Obs1}
\hat A= \sum_\alpha \alpha  E_A(\alpha),
\end{equation}
where $ E_A(\alpha)$ is projection onto the space ${\cal H}_A(\alpha)$ of eigenvectors for the eigenvalue $\alpha.$ 
For a pure state $\vert \psi \rangle \in {\cal H},$ the probability to get the outcome $A=\alpha$ is given by {\it the Born's rule}: 
\begin{equation}
\label{Obs2}
P_\psi(A=\alpha)= \Vert    E_A(\alpha) \psi \Vert^2=\langle  E_A(\alpha) \psi|\psi \rangle.
\end{equation}
A measurement with the  outcome $A= \alpha$ generates back-action onto system's state:
\begin{equation}
\label{Obs3}
\vert \psi \rangle \to \vert \psi \rangle_{A=\alpha}=  
 E_A(\alpha) \psi/ \Vert  E_A(\alpha) \psi \Vert.
\end{equation}
This is the projection postulate, one of the axioms of QM. In this general form it was formulated 
by L\"uders. Often one refers to (\ref{Obs3}) as the von Neumann projection postulate. However, 
von Neumann explores (\ref{Obs3}) only for observables represented by Hermitian operators with non-degenerate 
spectra \cite{VN}. In the case of degenerate spectra he considered more general state updates which in future led to quantum 
instrument theory.

Formula (\ref{Obs2}) implies the Born's rule for mixed states described as density operators,
\begin{equation}
\label{Obs2a}
P_\rho(A=\alpha)= \rm{Tr} \rho E_A(\alpha).
\end{equation}
A measurement with the  outcome $A= \alpha$ generates back-action onto system's state:
\begin{equation}
\label{Obs2b}
\rho \to \rho_\alpha=  E_A(\alpha) \rho E_A(\alpha)/ \rm{Tr} [E_A(\alpha) \rho E_A(\alpha)].  
\end{equation}

\subsection{The Quantum-like Paradigm and its Applications}

The QL paradigm has generated a rich literature across disciplines. In psychology, researchers have modeled decision-making anomalies using quantum probability theory, showing how interference effects explain deviations from expected utility theory, showing 
conjunction adn disjunction effects.  In economics and finance, QL models have been applied to volatility clustering, investor sentiment, and market bubbles. In linguistics and information retrieval, quantum logic has offered new ways to represent semantic ambiguity and relevance.

This paradigm also interacts fruitfully with the rise of quantum information theory, which shifts attention from particles and energies to information, communication, and computation within a quantum framework. Concepts such as quantum entropy, entanglement-based correlations, and nonlocality have found analogs in social systems, where agents exchange, process, and act upon information in non-classical ways.

As such, QLM is not merely a scientific curiosity or a mathematical toy - it represents a broader epistemic shift in how we conceptualize uncertainty, interaction, and systemic dynamics. It forms part of what some scholars describe as {\it the  Quantum-like Revolution} in the human sciences, parallel to the ongoing quantum information revolution in physics and engineering known as {\it the Second Quantum Revolution} \cite{Zeit}. Together, they signal a transformation in both theoretical perspective and technological possibility.

\subsection{Classical Decision Theory Under Fire}

The movement toward QLM is further supported by a deep critique of classical decision theory developed in cognitive psychology, particularly in the seminal work of Daniel Kahneman and Amos Tversky. Their research demonstrated that actual human decision-making systematically violates the assumptions of Savage's Subjective Expected Utility (SEU) theory, which posits that rational agents assign probabilities to events and choose actions that maximize expected utility based on those probabilities.

In landmark experiments and their analysis Kahneman, Tversky, and Shafir introduced anomalies such as the conjunction fallacy and the disjunction effect, showing that individuals often make choices inconsistent with the basic axioms of classical probability theory
\cite{kahneman1979,kahneman1981,kahneman1986,tversky1992,shafir1992,shafir1993b}. For example, in the disjunction effect, people refuse to make a decision in a situation of uncertainty, even when the possible outcomes, once disambiguated, would lead to the same choice. This violates the sure-thing principle, a cornerstone of classical decision theory.

Building on this, Eldar Shafir and colleagues explored how context and framing could shift preferences in non-linear and non-classical ways. Their findings showed that preferences are constructed in the act of decision, and not merely revealed - pointing to a deeper incompatibility between the assumptions of classical probability theory and the dynamics of human reasoning.

QLMs, with their built-in contextuality, interference effects, and probabilistic superpositions, offer a compelling alternative to account for these cognitive phenomena. Rather than treating anomalies as irrational deviations from an ideal, QL modeling provides a new normative and descriptive framework - one that better fits empirical data and respects the complexity of human thought.

\subsection{ Social Laser as a Consequence of the Quantum-like Turn}

Within this broader QL framework, SLT emerges not as a metaphorical flourish, but as a rigorous and innovative application of QL modeling to social dynamics. By extending the quantum conceptual toolkit from individual cognition to collective excitation and coherence, SLT suggests that mass social phenomena - such as protests, revolutions, or ideological mobilizations - can be understood in terms of quantized informational energy, population inversion, and stimulated emission of coordinated action.

This theory represents a natural evolution of the QL paradigm, coupling it with insights from quantum optics and embedding it within a socio-technological context dominated by informational amplification. In doing so, it offers a conceptual bridge between the quantum information revolution and the quantum-like revolution in social theory.

Where quantum computing harnesses coherent quantum states for computation, the Social Laser models how coherent social states can emerge from informational environments saturated with emotional and cognitive stimuli. It captures the discontinuous, collective shifts in human behavior that defy classical prediction and gives them a formal dynamical structure grounded in the logic of resonance, coherence, and critical thresholds.

Thus, SLT is not an isolated conceptual novelty, but a direct and compelling consequence of the QLM project. It exemplifies how quantum-inspired methods can illuminate not only individual cognition but the deeply collective, affective, and informational dynamics that shape modern society.

\section{Basics of Social Laser Theory}

The concept of SLT, introduced by Andrei Khrennikov, provides a novel framework for understanding how collective social action can be triggered and amplified through mechanisms reminiscent of quantum physics — particularly the functioning of a physical laser. In SLT, individuals within a society are treated as {\it social atoms} that possess quantized levels of {\it social energy.}
In SLT  an ensemble of social atoms forms {\it the gain medium} of lasing. 
These atoms interact within {\it a quantum information field} - a collective field of meaning, emotion, and perception shaped by the flow of information in society.

A key feature of SLT is the process of {\it social pumping}, which involves the transmission of emotionally charged, resonant information through mass communication channels. When a sufficient portion of the population is pumped into an excited social energy state by this information, and when the information is sufficiently coherent, social coherence can emerge.
This is the stage of {\it population inversion,} more than 50\% of population involved in the process of social lasing becomes 
 socially excited. (Initially the majority of social atoms are being passive, unexcited - by physical terminology 
 in the ground state.)   At this point, {\it a stimulated emission} of behavior may occur — manifesting as protests, mass movements, or viral trends.
Stimulation in the form of say news is generated by mass-media and social networks.  In this context, the analogy with laser action is not merely metaphorical but is structured around core principles such as energy quantization, coherence, and population inversion.

In physics, stimulated emission occurs when a beam of coherent radiation (photons) is injected into a laser's gain medium, triggering a cascade of coherent radiation from excited atoms. An injected photon interacts with an excited atom, causing the atom to emit a new photon that has identical characteristics to the incoming one. With each photon-atom interaction, the number of coherent photons effectively doubles. After about 100 such interaction steps, a single photon can generate approximately one million coherent photons.

A similar process can occur in a \emph{social gain medium}, where quanta of social energy - called \emph{infons} - play a role analogous to photons in physical lasers.

Schematically, both physical and social lasing follow the same basic mechanism:
\begin{enumerate}
    \item Energy pumping (physical or social).
    \item Population inversion.
    \item Stimulated emission.
\end{enumerate}

In section \ref{analogy}, we explore the analogy between the terminology and mechanisms of physical and social lasers in more detail.

\vspace{1em}

It is also worth noting that, aside from artificially engineered physical lasers, there are naturally occurring lasers in the universe - known as astrophysical or cosmic lasers (in fact, masers, see section~\ref{cosmic}). These typically involve a star surrounded by a gas cloud. The star acts as the energy-pumping source, while the gas cloud serves as the gain medium. Periodically, the gas cloud emits coherent radiation in pulses. However, unlike in typical lasers, these pulses result from spontaneous rather than stimulated emission (see section \ref{emissionf} on spontaneous vs. stimulated emissions). Roughly speaking, a single spontaneously emitted photon can trigger a cascade that generates a pulse of coherent photons identical to the initial one.

Spontaneous emission also occurs in engineered physical lasers. However, in such systems it is suppressed by an additional component—the resonator (often an optical cavity). The resonator not only prevents uncontrolled spontaneous cascades but also amplifies the laser beam and improves its coherence.

Similarly, a social laser should be equipped with a resonator to prevent spontaneous bursts of social energy, see article  
\cite{laser7} for details. Here, we simply note that such a component not only suppresses spontaneous cascades but also amplifies the beam of social energy and enhances its coherence.

\subsection{Correspondence Between Physical and Social Lasers}
\label{analogy}

In article \cite{laser6} Khrennikov, Toffano, and Dubois  draw a structural analogy between physical laser systems and collective human behavior by mapping components from quantum optics to elements in social dynamics.

\begin{table}[h!]
\centering
\begin{tabular}{ll}
\toprule
\textbf{Physical Laser Concept} & \textbf{Social/Information Laser Analogue} \\
\midrule
Atoms & Individuals (social atoms) \\
Energy levels & Mental or motivational states \\
Pumping (energy input) & Information bombardment (e.g., media, propaganda) \\
Stimulated emission & Triggered action or behavior \\
Laser radiation & Collective synchronized behavior or social action \\
\bottomrule
\end{tabular}
\caption{Analogy between components of physical and social lasers}
\end{table}

\section*{Key Concepts}

\begin{itemize}
  \item \textbf{Information Pumping}: Analogous to energy pumping in lasers, individuals are exposed to intensive, resonant information streams (e.g., news, social media), elevating their motivational or mental states.

  \item \textbf{Population Inversion}: When a critical mass of individuals reaches an excited mental state, society enters a preparatory phase for large-scale collective behavior.

  \item \textbf{Stimulated Emission in Social Contexts}: A triggering event (e.g., a protest, viral post) can induce a synchronized response across the population.

  \item \textbf{Coherence and Amplification}: As in laser physics, the social response can be coherent (well-aligned) and amplified, leading to large-scale actions from relatively small stimuli.
\end{itemize}

\subsection{Social Energy}

Social energy is a foundational concept in SLT, serving as the analog of physical energy in human societies. It represents the latent capacity of individuals or groups to act, mobilize, or respond to external stimuli. This energy is not merely metaphorical - it is postulated to have a quantifiable, dynamical character similar to energy in quantum systems. Social energy may be accumulated over time through various socio-cultural, economic, or psychological processes - such as political dissatisfaction, collective identity formation, or ideological alignment.

Importantly, social energy is not constantly active. Much like electrons in atoms, individuals in society can exist in ``low-energy'' (passive) or ``high-energy'' (activated) states. Social energy manifests visibly when mobilization occurs, e.g., in color revolutions, protests, or rapid shifts in public opinion. In the laser metaphor, this corresponds to the emission of a ``social signal'' or coordinated action. Hence, social energy serves both as a reservoir of potentiality and a mechanism of transformation in large-scale social dynamics.

See section \ref{energy} for further discussion on the notion of social (mental, psychic) energy, including the historical perspective. 

\subsection{Social Atom with Discrete Levels of Social Energy}

Since long time scientists tried to extend the idea of  Greek philosophers Leucippus and Democritus from the physical to 
mental world. Such activity was stimulated by intensive development of statistical mechanics and thermodynamics and
the strong interest in applying their methodology and formalism to mental (psychic) and social phenomena 
(see section \ref{SA}  for further discussion, including the historical perspective). To the beginning of this century such activity was practically terminated. Development of SLT 
again ignited interest to  it,  but in the quantum framework. 
     
The {\it social atom} is the basic unit of SLT, analogous to the physical atom in quantum mechanics. It represents an individual agent or actor in society, endowed with discrete internal states characterized by varying levels of social energy. These states can be understood as different modes of engagement or motivation — ranging from political apathy to full-scale activism.

The social atom is modeled as possessing a finite and quantized energy spectrum. Just as electrons occupy discrete orbitals, the individual in the social laser model is seen as jumping between distinct states rather than evolving continuously. These jumps are not deterministic but probabilistic, echoing the quantum-mechanical formalism.

Furthermore, a collection of such atoms may collectively exhibit phenomena like coherence and entanglement, where large numbers of individuals align their internal states in response to shared informational stimuli. This coherence is essential for the laser-like amplification process that characterizes mass social mobilization.

In SLT, quantization of social energy means that individuals do not transition between emotional, motivational, or behavioral states in a smooth, gradual fashion. Instead, these transitions occur in discrete steps, reflecting the quantized nature of internal states within the social atom. For instance, an individual might jump from a state of mild concern to intense protest - readiness without passing through intermediate stages.

This assumption borrows directly from quantum mechanics, where particles such as electrons can only occupy specific energy levels. Analogously, social atoms can only exist in defined configurations of internal motivation or energy. These quantized states are key to enabling the ``stimulated emission'' process central to the social laser metaphor: once a sufficient number of atoms are excited to a higher energy level, a coherent stimulus can trigger a mass transition, releasing social energy in a powerful, synchronized burst—akin to a laser pulse. This mechanism explains the suddenness and intensity of certain social phenomena, such as viral movements, uprisings, or ideological shifts, which seem to emerge spontaneously once a critical threshold is crossed.

Typically in QLM we consider two level social atoms with social energy levels $E_0$ and $E_1$ corresponding to the non-exited and excited 
states, denoted by the symbols $|E_0 \rangle$ and $|E_1 \rangle.$  Such an atom can absorb and emit only one quantum of social energy
\begin{equation}
\label{Q}
\Delta E= E_1- E_0.
\end{equation}   
This is the social version of the famous Bohr's rule for coupling of the energy levels of (two level) atom with the energy spectrum of absorbed and emitted radiation. 
 
The two level model is oversimplified, but it reflects the main features of the lasing scheme. We remark that in physics to construct a laser on two level atoms is impossible, due to the coincidence of Einstein coefficients, 
\begin{equation}
\label{B12}
B_{12}=B_{21},
\end{equation}
giving the rates of absorption and spontaneous emission  of energy quanta - photons, for an atom in the electromagnetic field. To build the real physical laser,  one should use atoms with minimum three energy levels, $E_0 < E_1 <E_2,$ but in practice one works with four level atoms, 
$E_0 < E_1 <E_2 < E_4,$ with the special constraints on the energy spectrum, $\Delta_{ij}= E_i-E_j.$ SLT for four level social atoms 
was schematically presented in \cite{laser2,laser5}, but it wasn't developed further. But in SLT there is no reason to assume constraint  (\ref{B12}). Therefore a social laser based on the gain medium consisting of two level social atoms can be created; moreover,  it seems that the majority of real social lasers, as e.g. used for color revolutions, are two level lasers.  

In physics the consideration of models with  atoms having of the discrete energy levels, $E_0 < E_1< ... <E_n,$ is based on the 
Fermi-Dirac statistics (section \ref{Istat}) for electrons; each energy level can be occupied by only one electron with the concrete value of spin. However, in a mathematical model by starting directly with $n$-levels system with states 
$|E_0\rangle, ..., |E_n\rangle$  we need  not  refer to  Fermi-Dirac statistics. 

As we have already mentioned electron's spin, this is the good place to point to the possibility for further development SLT    
 by introducing a kind of social spin (mental spin); the first step in this direction was done in \cite{laser7,laser9}, but such SLT framework wasn't developed further.  

We stress that in SLT the discreteness of the energy levels of social atoms is behavioral by its nature. But it may happen that 
such behavioral discreteness has neurophysiological roots. This is the interesting problem for furher studies directed to coupling of SLT with neurophysiology.    

\subsection{Information Field with Quanta of Information - Infons}
\label{infons}

The quantum information field (or social field) in SLT represents the environment that permeates society, transmitting information much like electromagnetic fields transmit light. However, rather than being continuous, this field is composed of discrete units or ``quanta of information'', analogous to photons in quantum electrodynamics. Therefore we highlight that the information field is quantum.  

These quanta of information are the basic carriers of communicative content—memes, slogans, symbols, or narratives that can stimulate or destabilize the internal states of social atoms. A quantum information field has properties such as coherence, polarization, and entanglement, and can affect large populations if the information is sufficiently resonant or synchronized with existing social energy levels.

For example, in times of political unrest, a single powerful message or symbolic act (a viral video, a protest image, a slogan) can resonate across the population, inducing synchronized reactions. This effect is similar to a coherent laser beam triggering emissions from excited atoms. Thus, the quantum information field not only transmits information but also shapes and amplifies collective behavior.
Qualitatively this field can be coupled (at least indirectly) to the notion of discursive frame,

\medskip

\textbf{Definition 1.} A \emph{discursive frame} is a structured narrative or interpretative schema that shapes how information is understood, categorized, and emotionally evaluated. It organizes facts and values into coherent storylines, influencing perception and behavior.

\medskip

Information quanta are called {\it infons.}  Infons mediate interactions within the quantum information field, specifically in the social context; infons are excitations of the quantum information field, similarly as photons are excitations of the quantum electromagnetic field. The term is coined by analogy to other fundamental quanta in physics (like photons, phonons, gluons, or gravitons) but applied to the realm of social communication and cognition.

Infons are conceived as the carriers of discrete informational impulses (quanta of social energy), such as emotionally charged messages, ideological cues, or directives for coordinated action. Importantly, infons are not classical messages or deterministic instructions; they possess a QL character, meaning that their effects depend on the contextual and probabilistic dynamics of the receiving system. A population exposed to the same infons may not all respond identically, but when a critical mass is in an excited state, these quanta can trigger large-scale coherence and stimulated emission—culminating in social mobilization.
Infons thus function as the essential triggers of the social laser mechanism: discrete, resonant, and capable of amplifying latent social energy into organized, collective action. We propose the following formal definition of an infon:

\medskip

\textbf{Definition 2.} An \emph{infon} is a theoretical quantum-like unit of social information. It represents a discrete, transferable package of social energy in coupling with meaning or intent capable of interacting with the cognitive states of individuals, social atoms.

\medskip

We repeat that infons are the information-theoretic analogues of photons in physical laser theory. They propagate within the \emph{quantum information field}, transmitting semantic impulses that may trigger mental excitation or alignment in social agents.

Each infon can induce a probabilistic change in the state of a social atom, depending on the resonance between the infon's content and the energy level structure of the recipient. 
To be absorbed or emitted by a social atom, the amount of social energy carried by infon should match the energy gap in this atom, 
\begin{equation}
\label{label}
\Delta = E.
\end{equation}.
If an infon carries the amount of social energy $E$ which isn't equal to the energy gap $\Delta,$  then a social atom wouldn't react to such information. Of course, in reality we don't have precise equalities. In physics, instead of equality (\ref{label}), we have 
a Gaussian distribution of energies of photons with mean value $\Delta.$ 

If $E <<\Delta$ or $E>>\Delta,$ then the probability that a social atom would interact with such infon is close to zero. 
People ignore  media messages carrying too low or too high amount of social energy 
comparing with their own energy spectrum. And different psychological reactions are behind these two behaviors:   
if, for some media message $C$ its social energy is very low comparing with the spectral gap in a social atom $a,$  $E <<\Delta,$ then 
$a$  simply `doesn't see' $C,$ if  $E>>\Delta,$ then $a$ would definitely 
pay attention to $C,$ but in such a situation $a$ should put in charge a mechanism of defense against psychic disturbance by  such high energy infons.   

This is the good place to  mention {\it phonons}, the quanta of vibrations, e.g., in a crystal, excitations of the field of mechanical  fluctuations.
While photons still match with our heuristic image of a particles (at least up to some degree), phonons are really 
immaterial entities carrying information about the relative fluctuations of atoms. So, phonons are not so different from infons. 
Both are mathematical entities useful for modeling of, respectively, mechanical and social vibrations. 

In fact, in physics characterization of quantum field excitations is not reduced to energy. For example,  a photon, an excitation of the quantum electromagnetic field,  also has polarization and spatial characteristics.   
Generally photon's state $|E \alpha\rangle$ is characterized by the parameters $E$=energy  and $\alpha$=(polarization, temporal and spatial extensions).  

Excitations of social-information field, infons,  can also have some characteristics supplementary to social energy and related to media message's content. In previous articles we called them {\it quasi-colors} of excitations (infons). This terminology was associated with coloring of energy in physics, light frequencies in the visible spectrum  correspond to colors. We recall that photon's energy $E$ and frequency $\nu$ are connected by the Einstein's formula  
\begin{equation}
\label{einstein}
E= h \nu,
\end{equation}
where $h$ is the Planck constant. This motivated the term `a quasi-color' for supplementary variables describing infon's state. 
However, this terminology might be misleading and doesn't match terms used in cognitive and social sciences.   
It seems that the widely used term `cognitive marker' match better SLT as a theory directed to cognitive and social phenomena,  

\medskip

\textbf{Definition 3.} A \emph{cognitive marker} is a semantically charged element embedded in a communicative act---such as a word, symbol, headline, tone, or visual cue---that triggers automatic interpretative or emotional responses by activating culturally or psychologically encoded meaning patterns.  These markers guide interpretation, frame meaning, and facilitate associative processing by activating pre-existing mental schemas, ideological frameworks, or affective dispositions.
 the receiver.

\medskip

In SLT, cognitive markers serve as \emph{semantic resonance units}. Their repetition and circulation help align individual mental states into collective modes, facilitating phase coherence necessary for social lasing. The use of the term `cognitive marker' is supported by the following arguments: 
\begin{itemize}
\item It is neutral and scientific—widely used in psychology, cognitive science, and linguistics.
\item It emphasizes the brain's role in filtering and reacting to information.
\item It suggests automatic processing: people often respond to cognitive markers without conscious reflection.
\end{itemize}

So, infon's state can be encoded as   $|E \alpha\rangle,$ where $E$ and $\alpha$ are social energy  and  cognitive marker  respectively.  Social atoms are humans who in some context  don't analyze content of media messages received from mass-media or social networks carefully; they operate with labels assigned to contents of media messages 
Such labels can be related to 
\begin{itemize}
\item Keywords - ``corruption'', ``war'', ``vaccination'', ``democracy'', ``dictatorship'', ``Trump'', ``Putin'',... .
\item Frames – In media studies, refers to the angle or perspective used to present the content.
\item Narratives – Broader storytelling structures or ideological themes.
\item Discursive labels – In discourse analysis, labels that shape interpretation and meaning.
\item Ideological markers – Signals of political, cultural, or moral positioning.
\item Cognitive cues – Signals meant to activate certain mental models or reactions.
\item Bias indicators – Signals of slant or partiality.
\item Agenda-setting elements – Terms that shape what the audience thinks is important.
\item Framing devices – Choices in wording or structure that guide interpretation.
\item Semantic primes – Fundamental linguistic elements that shape perception (from cognitive linguistics).
\end{itemize}
Typically a social atom reacts to such labels unconsciously that is without conscious determination. 
Such sort of information processing is closely coupled to popcorn thinking (see section \ref{popcorn} 
for further discussion). 

Why do people process information in such a way? It seems that one of the reasons for such behavior is information overload (see section \ref{IO}).   

\section*{Diagram: Cognitive Marker in the Social Laser Framework}

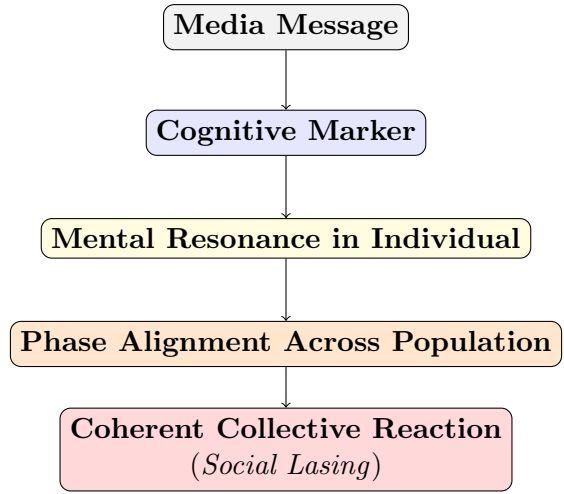
\begin{center}
\begin{tikzpicture}[node distance=1.4cm, every node/.style={align=center}]
  \node (msg) [draw, rectangle, rounded corners, fill=gray!10] {\textbf{Media Message}};
  \node (marker) [below of=msg, draw, rectangle, rounded corners, fill=blue!10] {\textbf{Cognitive Marker}};
  \node (resonance) [below of=marker, draw, rectangle, rounded corners, fill=yellow!15] {\textbf{Mental Resonance in Individual}};
  \node (alignment) [below of=resonance, draw, rectangle, rounded corners, fill=orange!20] {\textbf{Phase Alignment Across Population}};
  \node (lasing) [below of=alignment, draw, rectangle, rounded corners, fill=red!15] {\textbf{Coherent Collective Reaction} \\ (\emph{Social Lasing})};

  \draw[->] (msg) -- (marker);
  \draw[->] (marker) -- (resonance);
  \draw[->] (resonance) -- (alignment);
  \draw[->] (alignment) -- (lasing);
\end{tikzpicture}
\end{center}

\section*{Explanatory Note}

In SLT, {\it cognitive markers} operate as semantic quanta---discrete units of meaning that carry ideologically or emotionally charged content capable of stimulating synchronized mental responses. These markers are embedded within {\it  discursive frames}, which act as interpretative scaffolding that shape reception and coherence. When cognitive markers circulate repeatedly through communication channels, they function analogously to stimulated photons in a physical laser system: they align cognitive states across individuals, enabling the emergence of phase coherence. This alignment sets the stage for social lasing, manifesting as bursts of synchronized collective action or emotion.

\section{Quantum Versus Quantum-like Indistinguishability}
\label{INDIS}
      
Indistinguishability of systems is the main source of generation of quantum statistics which in turn plays the crucial role in physical and social lasing. Bose-Einstein statistics of photons and infons is behind the cascade processes of physical and social radiations.
			
\subsection{Indistinguishability in Quantum Physics}
\label{Istat}

In the classical treatment of statistical mechanics, particles are assumed to be distinguishable. This assumption leads to a paradoxical consequence: the entropy becomes non-extensive, growing faster than linearly with the number of particles. The resolution to this difficulty - and a fundamental aspect of quantum theory - is the recognition that elementary particles of the same type are {\it indistinguishable} in a much more profound sense.

In  quantum physics it is claimed that this indistinguishability is not merely a matter of lacking practical identifiers; it is a fundamental postulate of quantum theory. Two electrons, for instance, cannot be labeled even in principle: any attempt to describe a state that distinguishes them leads to a contradiction with the symmetry principles of quantum mechanics. This postulate has deep consequences for the statistical behavior of systems of many particles.

Let us now consider the derivation of quantum statistics from this principle. We examine a system of \( N \) identical particles. distributed among a set of quantum states labeled by \( i \), each with energy \( E_i \). The occupation number $n_i$ is the number of particles occupying the energy level $E_i.$  

Following Schr\"odinger's book \cite{SchT}, we consider energy not as an objective property of a 
particle, a property of an object, but as a measurable quantity that can be obtained as the result of measurement. Such treatment of energy is important for consideration of quantum systems. For classical systems, energy can treated objectively with the remark that 
this ``ontic value'' can be obtained from a measurement having sufficiently high precision. This convention can be used not only for 
energy, but for any observable. This is the good place to remark that Schr\"odinger's book \cite{SchT} is the most simple and strightforward introduction to classical and quantum statistical mechanics.    

In classical statistics, we count microstates by imagining that each particle carries a label. This counting 
leads to classical statistics of Maxwell-Boltzmann.

Quantum mechanics  treats particles as indistinguishable and it classifies identical particles into two types:
\begin{itemize}
  \item \textbf{Bosons}: particles whose total wavefunction is symmetric under particle exchange.
  \item \textbf{Fermions}: particles whose total wavefunction is antisymmetric under exchange.
\end{itemize}
This difference in symmetry leads to different rules for counting microstates.

The mean occupation number in the Bose-Einstein distribution is:
\begin{equation}
\label{S1}
\langle n_i \rangle = \frac{1}{e^{(E_i - \mu)/kT} - 1},
\end{equation}
where $k$ is the Boltzmann  constant, $T$ is temperature, and $\mu$ is a chemical  potential. For simplicity, we set $k=1$ (such system of units) and  $\mu=0$ and obtain the following formula:
\begin{equation}
\label{S2}
\langle n_i \rangle = \frac{1}{e^{E_i/T} - 1},
\end{equation}
  Due to the Pauli exclusion principle, no more than one fermion may occupy a given quantum state. 
The mean occupation number is:
\begin{equation}
\label{S3}
\langle n_i \rangle = \frac{1}{e^{(E_i - \mu)/kT} + 1},
\end{equation}
or again by setting $k=1$ and $\mu=0$ we get
\begin{equation}
\label{S4}
\langle n_i \rangle = \frac{1}{e^{E_i/T} + 1}
\end{equation}

\subsubsection*{Quantum Physics: Irreducible Indistinguishability}

Quantum  physics emphasizes that {\it the entire structure of quantum statistics arises from a simple yet profound constraint: particles are indistinguishable,} and the many-body wavefunction must reflect this. The indistinguishability leads not only to modified counting formulas, but also to physical phenomena without classical counterparts: the condensation of bosons into a single state (as in Bose-Einstein condensation), or the rigidity of the Fermi surface in metals.

In physics we may thus appreciate that quantum statistics is not an ad hoc modification of classical theory, but a deep consequence of the fundamental structure of quantum mechanics. In a sense, quantum statistics does not merely describe how particles \textit{behave} collectively—it reflects what they are.

\subsection{Quantum-like Modeling: Behavioral Indistinguishability}

We start this section by repeating the last statement of the previous section: {\it quantum statistics does not merely describe how particles \textit{behave} collectively—it reflects what they are.} This statement is about genuine quantum systems, as electrons or photons. It definitely cannot be applied to infons. Each infon, in the form of some media message, carries not only social energy, but also some content. Thus, two infons carrying the same amount of social energy can be distinguished by their content. Indistingushability of infons is generated by social atoms who don't analyze infon's content in very detail; they operate with labels assigned to contents, with cognitive markers (section \ref{infons}).

Operating with cognitive markers  is  related to such foundational issue as {\it indistinguishability of quantum 
systems}. Quantum theory assumes that two photons in the state $|E \alpha\rangle,$ where $\alpha$ denotes 
polarization and spatial direction variables,  are indistinguishable. Moreover, it is claimed that there are no hidden variables, additional to $E$ and $\alpha$  which might provide a possibility 
to distinguish two photons in the state $|E\alpha\rangle.$ Hence, in quantum physics indistinguishability has the fundamental character.
     
In QLM of the social-information field, infons are indistinguishable, up to social energy $E$ 
and the cognitive marker $\alpha.$ This is the important assumption behind SLT. However, there is one important difference between quantum and quantum-like indistinguishabilities.  The former is  genuine and irreducible and the later is relative 
to observers, to social atoms, and the context of interaction with the information field. In one context some social variables are important and they should be included in the cognitive markers $\alpha,$ in another context 
they do not play any role, so they are not included in $\alpha.$   But, we cannot deny their existence. For example, humans have names, but their names do not play any role in the process of social lasing in the form of mass protests. So, humans are indistinguishable w.r.t. to the name variable. The slogan `Black Lives Matter' is the integral cognitive marker which was crucial in the protests in USA. The concrete names of black people who experienced racism, discrimination, and racial inequality were hidden in this quasi-color.       

Thus indistinguishability in social laser theory is QL.  It is up to the characteristics determining   the process of lasing. These characteristics 
form the cognitive marker $\alpha.$ And infons' indistinguishability is up to social energy and this cognitive marker.    
For example, $\alpha$ can be `corruption', then infon's state is characterized by  social energy  and this cognitive marker. Hence, whole content of a media message is reduced to this $\alpha.$ Of course, a media message can carry a lot of additional information,  but in the process of this concrete (anti-corruption) social lasing it is ignored. 

{\it Can one derive quantum statistics on the basis of such behavioral indistinguishability?}

The answer is ``yes''. And  here we again refer to Schr\"odinger's book \cite{SchT}, where he  considered energy not as an objective property of a particle, a property of an object (ontic property), but as a measurable quantity that can be obtained 
as the result of measurement. Then, by using solely indistinguishability of systems, i.e. without any referring to quantum theory, particularly to (anti-)symmetry of wave functions, he derived the basic statistical formulas by pointing out that there are two logically possible cases: 
\begin{enumerate}
\item The occupations numbers $n_i$ can take any value, $n_i=0,1,2,.....$
\item The   occupations numbers $n_i$ can take only $q$ values, $n_i=0,1,2,..., q-1,$
where the number $q$ determines statistics.
\end{enumerate}
The first case corresponds to Bose-Einstein statistics.  In quantum physics only the case with   
$q=2,$ i.e., maximum one particle can occupy each energy state, is realized. In this way he obtained 
the Fermi-Dirac statistics. Statistics with $q>2$ are called parastatistics. Quantum physical systems 
following parastatistics  don't exist.

The Schr\"odinger's considerations are applicable to QL systems, e.g. to infons, see article \cite{laser1} and 
monograph \cite{laser5} for details. 

This is the good place to remark that at the present stage of development of QL science
 we don't have any reason to exclude parastatistics from consideration. There might exist cognitive-social contexts in which 
human's behavior follow parastatistics for some $q>2.$ 

\subsection{Quantum-like Modeling and Hidden Variables Problem}

The question of indistinguishability of QL systems discussed in the previous section is a part of the general problem of the interrelation of QLM and the hidden variables problem in quantum foundations.

Since the famous article of Einstein-Podolsky-Rosen (EPR) \cite{EPR}, the hidden variables problem has been disturbing quantum foundations. Foundational discussions led to numerous ``no-go'' theorems, as von Neumann, Bell, and Kochen-Specker. The general consensus is that hidden variables either do not exist or nonlocal. Nonlocality in the form of Einstein's spooky action at a distance, although widely accepted in quantum physics, is totally unacceptable in cognitive and social science. For cognition it leads to parapsychology and for social phenomena to esoteric rituals.  

In QL science hidden variables definitely exist and they are even not hidden - these are variables describing information processing be neuronal networks and in social science behavioral variables determining individuals psychic. Hence, it seems that employing of quantum theory for such macroscopic systems as brains and humans contradicts to quantum foundations. 

However, by stating  that  ``hidden variables don't exist'' one often forgets (or simply doesn't know) that this is the statement about non-contextual hidden variables. By considering context dependent hidden variables one can violate the Bell inequalities. And contextuality provides the resolution of the discussed foundational problem. Contextual hidden variables don't prevent employing the quantum theory. And cognitive and social behaviors are strongly context dependent. We note that contextuality is closely coupled to 
existence of incompatible variables. Such coupling  is the essence of Bohr's complementarity principle \cite{PL1}-\cite{PL4}. Finally, we highlight that the violation of the Bell inequalities is possible only in the presence of incompatible variables \cite{NL,NLa}.

\section{Popcorn Thinking and Social Laser Theory}
\label{popcorn} 

While not always precisely defined, popcorn thinking typically refers to:
\begin{itemize}
\item Rapid, scattered, spontaneous, and often superficial thoughts that ``pop up'' in the mind in response to stimuli.
\item A non-linear, fragmented style of consuming and processing information - especially prominent in digital media environments.
\item Often associated with information overload, scroll culture, or attention fragmentation.
\end{itemize}

\subsubsection*{How do Labeling Feed Popcorn Thinking?}

The labels used in digital and media content (e.g. video thumbnails, headlines, hashtags) serve as triggers in a popcorn-thinking environment. Here's how the connection plays out:

\begin{enumerate}
\item {\bf Cognitive Shortcutting.} Labels act as mental cues or priming tools. For example, seeing the word `scandal' in a title primes the brain for judgment or curiosity without deep processing. These labels encourage quick judgments or emotional reactions, consistent with popcorn-style cognition—rapid, surface-level, and reactive.
\item  {\it Fragmentation of Attention.}
When content is label-heavy (e.g., short-form videos, algorithmic feeds), it supports the fragmented, jumpy style of popcorn thinking.
The user is nudged to move rapidly from one labeled piece of content to the next, often without deeper reflection.
\item {\bf Ideological Reinforcement.}
Ideological labels or framing devices reinforce existing mental frames and echo chambers. In a popcorn-thinking mode, the brain tends to accept these at face value. The result is low-effort pattern recognition: ` I've seen this before, I know what to think,'' even if the context is different.
\end{enumerate}

We point to coupling between label-driven media content and popcorn thinking:
\begin{enumerate}
\item Labels/Frames in content $\implies$ Trigger Popcorn Thinking.
\item Headlines, tags, thumbnails	 $\implies$ 	Surface-level cognitive pops
\item Ideological markers, framing devices	 $\implies$ 	Reinforce shortcuts and emotional reactions
\item Algorithmic delivery of labeled content	 $\implies$	Encourages more popcorn-style engagement
\end{enumerate}
This coupling can shape public discourse, making it more reactive, polarized, or emotionally charged.

\subsubsection*{Analytical Framing}

By  analyzing this issue academically or conceptually, one might describe it like this:

In today's media ecosystem (basically digital), the proliferation of labels - be they algorithmic tags, ideological frames, or discursive shortcuts - interacts with a mode of engagement (social energy) that might be termed ``popcorn thinking.'' This mode is characterized by rapid, affect-driven responses to content fragments. Labels act as both cognitive triggers and ideological primers, reinforcing a fragmented attention economy where depth is often sacrificed for immediacy and familiarity.

\section{Spontaneous and Stimulated Absorption and Emission of Photons}

Atoms can interact with electromagnetic radiation through three fundamental processes: \textbf{absorption}, \textbf{spontaneous emission}, and \textbf{stimulated emission}. These processes are central to the quantum theory of light-matter interaction and play a crucial role in the functioning of lasers and other quantum devices.

\subsubsection*{Absorption}

When an atom in a lower energy state absorbs a photon whose energy matches the energy difference between two atomic levels, it transitions to a higher energy state. This process is known as \textbf{photon absorption} and occurs according to the energy conservation condition:
\begin{equation}
\label{F}
E_2 - E_1 = E_{\rm{photon}},
\end{equation}
where \( E_1 \) and \( E_2 \) are the atomic energy levels, \( h \) is Planck's constant, and $E_{\rm{photon}}$ is the energy of the absorbed photon.

\subsubsection*{Spontaneous Emission}

An atom in an excited state can return to a lower energy state by emitting a photon spontaneously. This process is called \textbf{spontaneous emission}. It occurs without any external influence and is a purely quantum phenomenon. The emitted photon has an energy corresponding to the difference between the two energy levels:
\begin{equation}
\label{F1}
E_{\rm{photon}} = E_2 - E_1,
\end{equation}

The direction and phase of the emitted photon are random in spontaneous emission. This process is responsible for phenomena such as fluorescence and the natural linewidth of atomic transitions.

\subsubsection*{Stimulated Emission}

In contrast to spontaneous emission, \textbf{stimulated emission} occurs when an atom in an excited state interacts with an incoming photon whose energy matches the energy difference between two levels. The interaction induces the atom to emit a second photon, which has the same energy, phase, polarization, and direction as the incoming photon. As a result, the electromagnetic field is amplified coherently.

Stimulated emission underlies the operation of lasers. The rate of stimulated emission increases with the intensity of the incident radiation, making it controllable and useful for coherent light generation.

\subsubsection*{Einstein Coefficients and Detailed Balance}

Albert Einstein introduced a quantitative description of these processes through the \emph{Einstein coefficients}:
\begin{itemize}
    \item \( B_{12} \): probability of absorption.
    \item \( A_{21} \): probability of spontaneous emission.
    \item \( B_{21} \): probability of stimulated emission.
\end{itemize}
By applying the principle of \emph{detailed balance} in thermal equilibrium, Einstein derived relationships among these coefficients, ensuring consistency with Planck's law of blackbody radiation.

In summary, photon absorption and emission processes govern the exchange of energy between atoms and radiation:
\begin{itemize}
    \item \emph{Absorption} excites the atom by destroying a photon.
    \item \emph{Spontaneous emission} de-excites the atom by randomly emitting a photon.
    \item \emph{Stimulated emission} induces coherent photon amplification.
\end{itemize}These processes form the foundation for understanding laser physics, quantum optics, and many applications in modern photonics and quantum technologies.

\subsubsection*{Remark on the Equality of \( B_{12} \) and \( B_{21} \) and the Impossibility of Two-Level Lasers}

For a two-level atom interacting with electromagnetic radiation, it follows from the microscopic reversibility of the light-matter interaction that the coefficients for absorption and stimulated emission are equal:
\[
B_{12} = B_{21}.
\]
This equality reflects the fundamental symmetry of absorption and stimulated emission processes under time-reversal. It means that, in thermal equilibrium, the rates of photon absorption and stimulated emission are exactly balanced (assuming equal populations), provided the radiation field remains unchanged.

As a consequence, it is \textbf{impossible to achieve lasing in a system consisting of only two atomic levels interacting with an electromagnetic field}. Even with population inversion (more atoms in the excited state than in the ground state), the absorption and stimulated emission rates remain equal because they are both proportional to the same Einstein coefficient and the same spectral energy density of the radiation field.

This result is often referred to as the \emph{two-level no-go theorem for lasers}: a laser cannot be built using a simple two-level atom interacting solely with electromagnetic radiation, because the net gain cannot overcome losses in such a system.

\textbf{Multi-Level Systems:} In practice, most lasers use at least three or four energy levels to achieve population inversion and suppress unwanted absorption. Intermediate levels serve as auxiliary states to facilitate pumping and selective relaxation.
    
\subsubsection*{Beyond the No-Go Theorem: Two-Level-Like Lasers in Alternative Systems}

While electromagnetic-field-based lasing is impossible in a true two-level atomic system, certain physical systems can still exhibit laser-like behavior involving effective two-level transitions under special conditions:
\begin{itemize}
    \item \textbf{Non-Radiative Pumping and Non-Equilibrium Systems:} In some systems, non-radiative processes (e.g., electrical injection or phonon-mediated transitions) can decouple the rates of absorption and stimulated emission, effectively bypassing the constraints of the two-level no-go theorem.
    
    \item \textbf{Quantum Coherence Effects:} Systems utilizing strong quantum coherence and interference, such as \emph{electromagnetically induced transparency} (EIT) or \emph{coherent population trapping}, can enable amplification without violating fundamental principles. These involve coherent manipulation of atomic superpositions beyond the naive two-level approximation.
    
    \item \textbf{Spin-Based Lasers and Masers:} In systems where the two levels correspond to spin states rather than electronic levels, lasing-like effects can occur via magnetic resonance or spin injection. These systems may exploit interactions beyond standard photon coupling, such as phonons or magnons.
\end{itemize}

Thus, although lasing is strictly forbidden in the simplest two-level atom–photon systems, many effective two-level-like systems can be engineered to exhibit laser-like amplification through the use of additional couplings, multi-step pumping, or strong coherence effects.

\section{Spontaneous and Stimulated Absorption and Emission of Infons}
\label{emissionf}

We recall that in the framework of SLT, societies can be modeled similarly to physical systems exhibiting laser-like amplification of collective actions. Here, the role of photons is played by discrete quanta of social energy, called \textbf{infons}. These infons are associated with specific information units capable of exciting social agents to higher levels of social energy.

\subsubsection*{Absorption of Infons}

In analogy with photon absorption by atoms, a social agent in a low-energy (passive) state can absorb an infon, thus transitioning to an excited (active) social state. This absorption process depends on the resonance between the infon's informational content, cognitive marker,  and the agent's cognitive and emotional predispositions.

\subsubsection*{Spontaneous Emission of Infons}

An excited social agent can spontaneously release an infon, returning to a lower social energy state. This \textbf{spontaneous emission of social energy} is random in terms of timing, direction, and informational content. It may manifest as isolated acts of expression, communication, or individual action without external stimulation.

\subsubsection*{Stimulated Emission of Infons}

The key mechanism for \textbf{social lasing} is the process of \textbf{stimulated emission of infons}. When an excited social agent encounters an incoming infon resonant with their current cognitive-emotional state, it can emit an additional infon identical in content, orientation, and social impact to the stimulating infon. This leads to a cascade-like amplification of social energy across the population, resulting in synchronized, coherent collective behavior.

This mechanism parallels the stimulated emission of photons in physical lasers and is central to explaining phenomena such as mass protests, viral social movements, or sudden bursts of coordinated collective actions.

\subsection*{Remark on the Two-Level No-Go Theorem for Social Lasing}

Similar to the physical case, the analogy with Einstein coefficients suggests that in a purely two-level social system, achieving stable social lasing via infon-induced interactions alone might be difficult. However, in SLT we don't have a no-go theorem, as in 
the physical laser theory.  

\subsection*{Beyond Two-Level Models: Realistic Social Lasing Mechanisms}

In realistic social systems, more complex mechanisms enable effective lasing-like behavior despite the two-level limitation:
\begin{itemize}
    \item \textbf{Multi-Level Social Structures:} Most societies exhibit multiple intermediate social states, such as levels of partial engagement, social roles, or ideological subgroups. These enable selective pumping and relaxation paths, creating effective population inversion among key groups.
    
    \item \textbf{External Non-Social Pumping:} Non-social stimuli---such as economic crises, political events, or disasters---can inject social energy non-radiatively, disrupting equilibrium and creating conditions for lasing-like bursts of social activity.
    
    \item \textbf{Cognitive Resonators and Amplifiers:} Social media algorithms, echo chambers, and targeted media strategies act as effective \emph{social resonators}, selectively amplifying certain infon frequencies while suppressing others, making lasing more likely even in seemingly two-level agent structures.
    
    \item \textbf{Collective Cognitive Effects:} Strong coherence in collective attention, emotional synchronization, and shared narratives can lock agents into tightly coupled dynamic states, effectively bypassing the two-level no-go limitation.
\end{itemize}

These mechanisms explain why large-scale coherent social phenomena, resembling lasing, are possible in complex societies despite the theoretical limitations of simplified models.

\section {The Role of the Optical Resonator in Physical Laser}

A fundamental component of any laser system is the \textbf{optical resonator}, commonly formed by two or more mirrors facing each other with a gain medium placed in between. The resonator plays a crucial role in two key aspects of laser operation:
\begin{enumerate}
    \item \textbf{Amplification of light} through stimulated emission.
    \item \textbf{Enhancement of coherence} of the emitted laser beam.
\end{enumerate}

\subsection{Amplification via Stimulated Emission}
Inside the optical cavity, light undergoes multiple reflections between the mirrors, repeatedly passing through the gain medium. Each pass triggers \textit{stimulated emission}, adding photons with the same phase, frequency, and direction to the beam.

The amplification of light intensity per pass through the gain medium can be expressed as:
\[
I_{\text{out}} = I_{\text{in}} e^{g L},
\]
where:
\begin{itemize}
    \item \(I_{\text{in}}\) and \(I_{\text{out}}\) are the input and output light intensities for one pass,
    \item \(g\) is the gain coefficient of the medium (per unit length),
    \item \(L\) is the length of the gain medium.
\end{itemize}

Over multiple round trips inside the cavity, the intensity grows exponentially until the gain balances the losses. This process produces a strong and intense laser beam. One of the mirrors is partially transmissive, allowing a portion of the amplified light to escape as the laser output.

\subsection{Enhancement of Coherence via Mode Selection}
The resonator acts as a frequency and mode selector, enhancing the \textbf{coherence} of the laser beam. Only specific optical modes satisfying the \textit{resonance condition} are strongly amplified:
\begin{equation}
\label{resonatorP}
2 L = m \lambda,
\end{equation}
where:
\begin{itemize}
    \item \(L\) is the optical cavity length,
    \item \(\lambda\) is the wavelength of the light,
    \item \(m\) is an integer representing the longitudinal mode number.
\end{itemize}

This condition ensures that only wavelengths for which an integer number of half-wavelengths fit within the cavity can persist. As a result, the laser achieves:
\begin{itemize}
    \item \textbf{High temporal coherence}: The output is nearly monochromatic with narrow spectral linewidth.
    \item \textbf{High spatial coherence}: The beam has a well-defined phase front and low divergence.
\end{itemize}

The table below summarizes the two main functions of the optical resonator:

\begin{center}
\begin{tabular}{|c|c|c|}
\hline
\textbf{Function} & \textbf{Mechanism} & \textbf{Effect on Laser Beam} \\
\hline
Amplification & Multiple round trips & Strong, high-intensity output \\
\hline
Coherence Enhancement & Mode selection  & High  coherence \\
\hline
\end{tabular}
\end{center}

\section{The Role of the Resonator in a Social Laser}

The \textbf{social laser theory} is a quantum-like approach to modeling collective social phenomena, drawing an analogy with physical lasers. In this framework, the \textbf{resonator} corresponds to social structures that enable the amplification and coherent organization of social energy.

Among these structures, \textbf{social networks}—especially in the form of \textbf{echo chambers}—play a pivotal role in two key processes:
\begin{enumerate}
    \item \textbf{Amplification of social energy} through repeated social interactions and reinforcement.
    \item \textbf{Enhancement of social coherence} by filtering and aligning opinions, narratives, and behaviors.
\end{enumerate}

\subsection{Amplification of Social Energy}
In a social laser, individuals (``social atoms'') can be excited to higher states of social energy by external influences such as media, propaganda, or events. Social networks act as resonators by repeatedly circulating information, thus reinforcing emotional and cognitive excitation.

Echo chambers, where individuals are mainly exposed to views and narratives that confirm their pre-existing beliefs, serve as \textit{high-gain social resonators}. Each interaction or exposure within such a network can increase the level of social excitation, analogous to stimulated emission in physical lasers.

This amplification process can be abstractly described as:
\[
E_{\text{out}} = E_{\text{in}} e^{g_s N},
\]
where:
\begin{itemize}
    \item \(E_{\text{in}}\) and \(E_{\text{out}}\) represent the initial and resulting levels of social energy,
    \item \(g_s\) is the social gain factor, reflecting the reinforcing power of the network,
    \item \(N\) is the effective number of reinforcing interactions or cycles within the network.
\end{itemize}

\subsection{Enhancement of Social Coherence}
Social resonators also act as coherence-enhancing structures. Echo chambers selectively amplify particular discourses or behavioral patterns, filtering out alternative views. This is analogous to the mode selection in physical laser cavities.

The condition for social coherence can be metaphorically represented as:
\begin{equation}
\label{resonatorS}
2L_s = m_s \lambda_s,
\end{equation}
where:
\begin{itemize}
    \item \(L_s\) is the effective ``length'' or reach of the social network,
    \item \(\lambda_s\) is the characteristic wavelength of a dominant social narrative or behavior,
    \item \(m_s\) is an integer representing the number of social ``resonance cycles'' or ideological reinforcements.
\end{itemize}

This condition ensures that only certain narratives, emotional frames, or action patterns—those fitting the social resonance structure—are strongly amplified, leading to:
\begin{itemize}
    \item \textbf{High cognitive coherence}: Shared narratives and beliefs across the group.
    \item \textbf{High behavioral coherence}: Coordinated and synchronized collective actions.
\end{itemize}

The table below summarizes the two main functions of the social resonator in a social laser:
\begin{center}
\begin{tabular}{|c|c|c|}
\hline
\textbf{Function} & \textbf{Mechanism} & \textbf{Effect on Collective Behavior} \\
\hline
Amplification & Repeated reinforcement in networks & Rapid increase in social energy \\
\hline
Coherence Enhancement & Alignment via social resonance & Unified and coordinated actions \\
\hline
\end{tabular}
\end{center}

\subsection{Deeper Interpretation of the Resonance Condition}

In SLT  we considered social resonance condition (\ref{resonatorS}). This condition was introduced as a metaphor to physical resonance condition  (\ref{resonatorP}). Can we proceed further to sign social network meaning to the parameters $L_s, m_s, \lambda_s?$ (see also section \ref{Tdis}). 

\subsubsection*{Effective Social Network Reach ($L_s$)}
This parameter represents the capacity of a social network to spread and retain information, including:
\begin{itemize}
    \item Size and density of the network,
    \item Communication efficiency,
    \item Influence dynamics,
    \item Amplification effects from media algorithms.
\end{itemize}

\subsubsection*{Characteristic Wavelength ($\lambda_s$)}
This reflects the scale or frequency of social narratives:
\begin{itemize}
    \item Small $\lambda_s$ corresponds to rapidly changing trends or viral memes.
    \item Large $\lambda_s$ corresponds to stable ideologies or long-term narratives.
\end{itemize}

\subsubsection*{Number of Resonance Cycles ($m_s$)}
This parameter describes the number of reinforcement loops a narrative undergoes within the social network:
\begin{itemize}
    \item Large $m_s$ indicates a strong echo-chamber effect.
    \item Higher $m_s$ corresponds to deeper ideological entrenchment.
\end{itemize}

\subsubsection*{Modeling Selective Amplification}
Equation \eqref{resonatorS} suggests that only certain narratives with wavelengths satisfying the resonance condition will be amplified within a society:
\begin{itemize}
    \item Narratives satisfying $2L_s \approx m_s \lambda_s$ are selectively amplified.
    \item Narratives not matching this condition tend to dissipate.
\end{itemize}

\subsection*{\textbf{Social Resonance Index}}
We define a social resonance index:
\begin{equation}
R_s = \frac{2L_s}{\lambda_s}
\end{equation}
Amplification occurs when $R_s \approx m_s$ (close to integer values).

\subsubsection*{Beyond Metaphor: Toward Quantitative Modeling}
This framework allows modeling:
\begin{itemize}
    \item Echo chamber formation and reinforcement dynamics.
    \item Viral spread of information.
    \item Mass mobilization and collective actions.
\end{itemize}

Applications:

\begin{itemize}
    \item Predicting social unrest and mobilization waves.
    \item Studying viral dynamics in social media.
    \item Modeling long-term ideological shifts.
\end{itemize}

\subsection{Towards a Mathematical Framework for Social Resonators}

SLT suggests that mass social movements can be understood via analogies with physical laser systems. One key concept is the \emph{social resonator}, analogous to an optical cavity, where a social network can support standing waves of dominant narratives or collective behaviors. This is captured by the resonance condition (\ref{resonatorS}):
As was pointed out, this condition remains largely metaphorical. Below, we explore possible mathematical formalizations and modeling paths.

\subsubsection{Spectral Graph Theory Approach}
Social networks can be represented as graphs $G=(V,E)$, where nodes represent individuals and edges represent interactions 
\cite{Chung}. The graph Laplacian $L$ captures diffusion and synchronization properties.

Eigenvalues $\lambda_i$ of $L$ correspond to characteristic frequencies of network modes. The \emph{algebraic connectivity} (second smallest eigenvalue) relates to the coherence of the network.

\textbf{Proposed Link:}
\begin{itemize}
    \item $\lambda_s$ could correspond to $\lambda_i^{-1/2}$ for selected $i$, reflecting coherent oscillatory modes.
    \item $L_s$ can be associated with the graph diameter $D$ or average shortest path length.
    \item Equation \eqref{resonatorS} then becomes:
    \[
    2 D \approx m_s \lambda_i^{-1/2},
    \]
    with $m_s$ capturing network reinforcement cycles.
\end{itemize}

\subsubsection{Wavelet and Spectral Filtering on Graphs}
\emph{Graph wavelets} allow localization of frequencies on networks. Using graph Fourier transforms, one can study narrative frequencies localized on communities or regions \cite{Barahona}.

\textbf{Approach:}
\begin{itemize}
    \item Model narrative spread as wave-like signal $f$ on graph.
    \item Decompose $f$ via graph wavelets at scale $s$, identifying $\lambda_s$ as dominant narrative scale.
    \item Social resonance then reflects alignment between $L_s$ and narrative scale $s$.
\end{itemize}

\subsection{Topological Data Analysis (TDA) Approach}
TDA \cite{Lum} techniques like \emph{persistent homology} capture multi-scale cycles in networks.

\textbf{Hypothesis:}
\begin{itemize}
    \item Number of persistent cycles $m_s$ corresponds to ideological reinforcement loops.
    \item Effective network scale $L_s$ can be linked to filtration thresholds at which these cycles appear.
    \item Equation \eqref{resonatorS} can thus be mapped onto birth-death diagrams in persistence barcodes.
\end{itemize}

\subsubsection{Agent-Based Modeling}
Simulations with agents exchanging narratives and influencing each other can explicitly implement the resonance condition.

\textbf{Suggested Model:}
\begin{itemize}
    \item Agents positioned on a network with given $L_s$.
    \item Narrative propagation modeled as wave-like reinforcement with decay.
    \item Measure whether reinforcement cycles $m_s$ and characteristic spread $\lambda_s$ satisfy Eq.\eqref{resonatorS} in simulated dynamics.
\end{itemize}

These three approaches---spectral graph theory, graph wavelets, and TDA---provide concrete ways to quantify and test the resonance condition \eqref{resonatorS} within mathematical social network models.

This represents a novel research direction unifying SLT with established mathematical tools, potentially leading to:
\begin{itemize}
    \item Identification of critical narrative frequencies enabling social coherence.
    \item Quantitative measures for social resonance thresholds.
    \item Predictive models for large-scale social mobilization.
\end{itemize}

\subsection{Distinction from Existing Notions}

While the concept of \emph{effective social network reach} ($L_s$) may seem related to classical measures such as \emph{effective diameter} and \emph{influence radius}, it is fundamentally different in both interpretation and intended use.

\subsection*{Effective Diameter}

In network theory, the \emph{effective diameter} of a graph refers to the $q$-th percentile of shortest path lengths between node pairs. A common choice is $q=0.9$, meaning the effective diameter is the smallest integer $d$ such that at least $90\%$ of all connected node pairs have shortest-path distance at most $d$. Mathematically,
\[
D_{\mathrm{eff}} = \min \left\{ d \, \bigg| \, \frac{|\{(u,v) \,:\, \mathrm{dist}(u,v) \leq d \}|}{|\{(u,v)\}|} \geq q \right\}.
\]
This metric reflects the compactness of the network and is widely used in small-world and Internet topology studies.

\subsection*{Influence Radius and Reachability}

The concepts of \emph{influence radius} and \emph{reachability} originate from studies of information diffusion, percolation, and epidemic models on networks. They are often defined in terms of how far an initial source of influence can propagate.

A formal approach involves modeling information spread as a probabilistic diffusion process on a graph $G = (V, E)$:
\begin{itemize}
    \item Let $p_{uv}$ denote the probability that node $u$ influences node $v$ (often derived from edge weights or fixed thresholds).
    \item The \emph{reachability set} $R(u, T)$ after $T$ time steps consists of nodes that can be influenced by $u$ within $T$ steps under this model.
\end{itemize}

The \emph{influence radius} can then be defined as:
\[
r_{\mathrm{inf}}(u, \varepsilon) = \min \left\{ r \, \bigg| \, \frac{|R(u, r)|}{|V|} \geq \varepsilon \right\},
\]
where $\varepsilon$ is the desired fraction of the network to be reached. This captures the minimal number of steps needed for node $u$ to influence at least a fraction $\varepsilon$ of the network.

Specific models where such notions are formalized include:
\begin{itemize}
    \item \textbf{Independent Cascade Model (ICM):} Influence spreads probabilistically along edges with given transmission probabilities.
    \item \textbf{Linear Threshold Model (LTM):} Nodes adopt a behavior once a weighted sum of their neighbors exceeds a threshold.
    \item \textbf{Percolation Theory:} Studies connected components under probabilistic edge activation.
\end{itemize}

These models offer rigorous mathematical tools for analyzing influence propagation and defining metrics such as influence radius.

\subsection*{Novelty of Effective Social Network Reach (\(L_s\))}

In contrast to both effective diameter and influence radius, the {\it effective reach} $L_s$  of a social network was invented by the author of this work. This notion determine to a functional feature of social networks coupled  to the phenomenon of \emph{social resonance}. We now discuss this notion in more details: 

\begin{itemize}
    \item $L_s$ expresses the characteristic of a social network indicating its ability to support and strengthen self-reinforcing cycles of narratives or behaviors.
    \item It is useful to describe resonance phenomena such that  the network successfully amplifies special frequencies of social narratives, similarly  to standing waves in physics, classical and quantum.
    \item The basic coupling  condition $2 L_s = m_s \lambda_s$
reflects matching between the structure of a network including its representation as graph  
and the main social wavelength $\lambda_s$, matching which allow for resonance strengthening.
\end{itemize}

The effective reach $L_s$  of a social network depends on both network topology and dynamic content spreading properties. 
In this sense $L_s$ is distinct from static structural or diffusion-based measures. The following table may be useful of a reader of this artcile:
\begin{center}
\begin{tabular}{@{}llll@{}}
\toprule
\textbf{Concept} & \textbf{Type} & \textbf{Meaning} & \textbf{Role in Modeling} \\ \midrule
Effective Diameter & Structural (global) & Typical shortest path lengths & Network compactness \\
Influence Radius / Reachability & Structural + Dynamic & Extent of influence propagation & Diffusion and spread dynamics \\
Effective Social Network Reach ($L_s$) & Functional (resonance) & Network's capacity for coherent resonance & Amplification and synchronization \\ \bottomrule
\end{tabular}
\end{center}

\section{Information Overload in Modern Society and Decision-Making}
\label{IO}

In today's hyper-connected world, individuals are exposed to an unprecedented volume of information. Every day, billions of data points  - news, articles, social media posts, videos, advertisements, and expert opinions — flow through an ever-expanding web of mass media, news agencies, and digital platforms of newmerous social networks. While access to information is widely considered a cornerstone of democratic societies and modern life, the sheer quantity and speed of information today have created a new and pressing problem: information overload. This section explores how the deluge of information affects individuals and societies, particularly in relation to cognitive processes and decision-making.

\subsubsection*{The Anatomy of Information Overload}

Information overload occurs when the volume of input to a system exceeds its processing capacity. For the human brain, this means being bombarded with more information than it can reasonably process, evaluate, and integrate. Herbert A. Simon, a pioneer in information science, noted that {\it ``a wealth of information creates a poverty of attention.'}' This insight is especially pertinent in the digital era, where cognitive resources are stretched thin across competing stimuli.

Mass media and news agencies contribute to this overload by operating on a 24/7 cycle that prioritizes speed over depth. In parallel, social networks amplify and fragment content through virality, personalization algorithms, and user engagement metrics. These mechanisms not only increase the quantity of information but also accelerate its dissemination, often at the expense of accuracy, coherence, or relevance.

\subsubsection*{Sources of Overload: Mass Media, News Agencies, and Social Networks}

\begin{itemize}
\item {\bf Mass Media.} Traditional media - television, radio, and newspapers - have evolved from being periodic and curated to near-continuous and reactive. The shift toward real-time reporting and multimedia storytelling means that individuals are now not only receiving headlines but also accompanying commentary, analysis, and cross-platform engagement. The need for ratings and ad revenue can further skew content toward sensationalism and emotional impact, rather than balanced information.
\item {\bf News Agencies.}
News agencies, once considered gatekeepers of verified, objective reporting, now face pressure from both market competition and digital immediacy. In the race to break news first, agencies often sacrifice investigative rigor for click-worthy content. Furthermore, syndicated news cycles can lead to repetition and redundancy, where similar stories are recycled across different platforms, contributing to cognitive fatigue and perceived urgency.
\item {\bf Social Networks.}
Perhaps the most significant accelerant of information overload is the rise of social media platforms such as Facebook, X (formerly Twitter), Instagram, and TikTok. These platforms blur the line between content creator and consumer, creating feedback loops that amplify particular themes or narratives. Algorithmic personalization tailors information to user preferences, often reinforcing cognitive biases and creating echo chambers. The consequence is not just more information, but more emotionally charged, fragmented, and context-deprived information.
\end{itemize}

\subsubsection*{Psychological and Cognitive Consequences.}

\begin{itemize}
\item {\bf Decision Paralysis.}
When faced with too many choices or conflicting information, individuals may experience decision paralysis—a state in which no choice is made due to the overwhelming complexity or fear of making the wrong choice. This is evident in everyday contexts such as consumer behavior, where users abandon online carts due to overabundance of options, but also in high-stakes arenas such as political participation, health choices, or financial planning.
\item {\bf Reduced Critical Thinking.}
Continuous exposure to rapid and emotionally charged information can erode critical thinking. The ability to evaluate sources, weigh evidence, and form independent judgments diminishes under pressure to respond quickly. This problem is compounded by the rise of misinformation, deepfakes, and manipulated content, which exploit overloaded minds less likely to scrutinize what they read or watch.
\item {\bf Emotional Fatigue and Anxiety.}
Overload doesn’t merely tax cognitive resources; it also induces emotional fatigue. Constantly consuming crisis-oriented or negative content - commonly referred to as ``doomscrolling'' - can lead to increased anxiety, helplessness, and even depression. This emotional toll further impairs rational judgment and may lead to impulsive or avoidance-based decisions.
\end{itemize}

\subsubsection*{Social and Behavioral Impacts}

\begin{itemize}
\item{\bf Polarization and Groupthink.}
When people are overwhelmed by conflicting data, they tend to default to heuristics, such as trusting familiar sources or aligning with group norms. This contributes to political polarization and groupthink, where communities reinforce their own beliefs and exclude dissenting voices. The structure of online networks exacerbates this by promoting content that aligns with users’ existing views, rather than challenging them.
\item {\bf Shortened Attention Spans.}
Studies have shown that average attention spans are decreasing, especially among younger populations. The constant stream of fast-moving content—optimized for likes, shares, and virality—conditions users to scan rather than reflect. This habit of superficial engagement makes it harder to retain complex information or make deliberative decisions.
\item {\bf The Rise of Manipulative Technologies.}
In an environment of cognitive overload, individuals become more vulnerable to manipulative technologies, such as behavioral nudges, algorithmic targeting, and persuasive design. These tools exploit limited attention and impulsive behavior to steer choices—whether in shopping, voting, or belief formation—in ways that are not always transparent or aligned with the user’s long-term interests.
\end{itemize}

\subsubsection*{Coping Mechanisms and Potential Interventions}
\begin{itemize}
\item {\bf  Media Literacy.}
Promoting media literacy—the ability to critically assess, interpret, and question media content—is a fundamental strategy to counter information overload. Educational systems, civil society organizations, and digital platforms must work together to equip individuals with tools to recognize bias, check sources, and manage their consumption habits.
\item {\bf Algorithmic Transparency.}
Platforms that use personalization algorithms should be more transparent about how content is selected and presented. Giving users control over algorithmic curation—such as chronological feeds or content filters—can mitigate the feeling of being overwhelmed and manipulated.
\item {\bf Digital Hygiene and Mindful Consumption.}
Behavioral techniques, such as setting time limits on apps, engaging in “digital detoxes,” or using curated newsletters instead of unfiltered feeds, can help individuals take back control of their information environment. Mindfulness practices may also enhance focus and reduce the anxiety associated with information overload.
\end{itemize}

{\bf Summary:} While access to information is a hallmark of progress, the unregulated abundance of it in modern society presents serious challenges to human cognition and decision-making. Mass media, news agencies, and especially social networks contribute to a landscape where attention is fragmented, judgment is impaired, and the quality of public discourse declines. Navigating this new terrain requires a multi-pronged effort—from education and technology to policy and individual behavior. Only by recognizing the complexity of information overload can we hope to restore clarity, agency, and rationality in the way we make decisions in the digital age

\section{Information Overload and Social Laser Theory}

In this theoretical light, information overload assumes a dual and paradoxical role.

\subsubsection*{Overload as a Precursor to Coherence.}
On one hand, information overload might appear antithetical to coherence. A population saturated with inconsistent, contradictory, or fragmentary information may become paralyzed or disengaged. However, SLT suggests that when information is emotionally aligned and repetitive across platforms, even if abundant, it can serve as a mechanism of coherent pumping. Overload is not just a byproduct of modern media systems; it is part of the necessary background condition in which resonant information—clear, emotionally engaging, and synchronized across multiple channels—can stand out and drive collective excitation.

This idea resonates with how advertising, propaganda, and political messaging often work: among the chaos of competing messages, it is the ones that strike a chord—emotionally, ideologically, or narratively—that become focal points of mass attention. SLT posits that such resonance can lead to a population inversion, where more individuals are in an excited social state than in a passive one, thus enabling the emergence of collective action.

Overload as a Catalyst for Quantum-like Social Transitions
From the SLT perspective, individuals under conditions of constant information bombardment may experience a form of social decoherence—a fragmentation of individual rationality and decision-making capacity. Yet ironically, this same decoherence can serve as the groundwork for QL phase transitions in social behavior. The overwhelmed cognitive state of individuals allows emotionally charged memes or narratives to act like infons - discrete quanta of social meaning that propagate through the social field and can trigger a synchronized behavioral response.

Thus, information overload functions as both environmental noise and a medium of amplification. While it may diminish individual deliberation, it enhances the susceptibility of the social field to resonance phenomena. In this way, SLT offers a sophisticated lens through which to understand contemporary events such as the rapid spread of political polarization, mass protests, viral misinformation, and sudden shifts in public sentiment.

Toward a Coherent Understanding
Ultimately, the relationship between information overload and collective behavior is more than just a problem of individual cognition; it is a systemic feature of modern information ecosystems. SLT helps explain how information overload, far from merely numbing populations, may set the stage for organized bursts of social energy, akin to the emission of light from a physical laser. The “light” in this case is collective behavior—emergent, coherent, and sometimes explosive.

Understanding this dynamic underscores the need for deeper media literacy, ethical design of communication technologies, and socio-political awareness of how information systems interact with mass psychology. As SLT makes clear, modern societies are not only saturated with data but also finely tuned resonators in which coherent action can be stimulated—sometimes with transformative, sometimes with destabilizing effects.

\section{Mean Field Approach to Social Laser}

In SLT, the mean-field theory approach helps to analytically describe the threshold conditions and dynamic evolution of social coherence.
We consider a population of $N$ social agents (individuals), each of which can be in one of two QL states:
\begin{itemize}
  \item Ground state $|0\rangle$: socially inactive
  \item Excited state $|1\rangle$: socially activated (ready to emit social action)
\end{itemize}
The population inversion parameter is:
\begin{equation}
  D = \langle \hat{\sigma}_z \rangle = \frac{1}{N} \sum_{j=1}^N \langle \hat{\sigma}_z^{(j)} \rangle
\end{equation}
where $\hat{\sigma}_z$ is the Pauli $z$-matrix, representing the difference in occupation between excited and ground states.

\subsubsection*{Mean Field Hamiltonian}

Following the quantum optical analogy, the mean-field Hamiltonian for the social laser reads:
\begin{equation}
  \hat{H} = \hbar \omega_0 \hat{a}^\dagger \hat{a} + \frac{1}{2} \hbar \omega_s \sum_{j=1}^N \hat{\sigma}_z^{(j)} + \hbar g \sum_{j=1}^N \left(\hat{a}^\dagger \hat{\sigma}_-^{(j)} + \hat{a} \hat{\sigma}_+^{(j)}\right)
\end{equation}
where:
\begin{itemize}
  \item $\omega_0$ is the frequency of the collective field (quantum information field)
  \item $\omega_s$ is the social excitation energy
  \item $g$ is the coupling constant
  \item $\hat{a}, \hat{a}^\dagger$ are annihilation and creation operators for the collective mode (social information field)
  \item $\hat{\sigma}_\pm$ are the raising and lowering operators for agents
\end{itemize}

\subsubsection*{Heisenberg-Langevin Equations}

In the mean field approximation, the dynamics are governed by the Heisenberg-Langevin equations:
\begin{align}
  \frac{d\hat{a}}{dt} &= -\left(\kappa + i \omega_0\right)\hat{a} + i g N \langle \hat{\sigma}_- \rangle + \hat{F}_a(t) \\
  \frac{d\hat{\sigma}_-}{dt} &= -\left(\gamma_2 + i \omega_s\right)\hat{\sigma}_- + i g \hat{a} \hat{\sigma}_z + \hat{F}_\sigma(t) \\
  \frac{d\hat{\sigma}_z}{dt} &= -\gamma_1 (\hat{\sigma}_z - D_0) - 2i g (\hat{a}^\dagger \hat{\sigma}_- - \hat{a} \hat{\sigma}_+)
\end{align}
where:
\begin{itemize}
  \item $\kappa$ is the decay rate of the collective field
  \item $\gamma_1, \gamma_2$ are longitudinal and transverse relaxation rates for social agents
  \item $D_0$ is the pump-induced inversion
  \item $\hat{F}_a(t), \hat{F}_\sigma(t)$ are noise operators
\end{itemize}

\subsubsection*{Threshold Condition}

The system undergoes a phase transition (social lasing) when the gain from population inversion exceeds losses. The threshold condition is derived by setting the time derivatives to zero and solving for steady-state. The critical inversion is:
\begin{equation}
  D_c = \frac{\kappa \gamma_2}{g^2 N}
\end{equation}

When $D_0 > D_c$, spontaneous coherence emerges in the social information field:
\begin{equation}
  \langle \hat{a} \rangle \neq 0 \Rightarrow \text{coherent social action}
\end{equation}

{\bf Summary:} The mean-field theory of the social laser, as presented by Alodjants et al. \cite{laser12}, formalizes the emergence of large-scale collective action through analogies with quantum optics. The appearance of coherent macroscopic fields in social systems can be viewed as a result of population inversion and collective coupling, leading to a dynamical phase transition. This framework provides both qualitative insight and quantitative tools to understand modern social phenomena.

\section{Random Social Laser}

Article \cite{laser13} explores the analogy between random lasers (RLs) in physics and complex social processes, focusing on how RLs can serve as simulators for social phenomena such as spontaneous collective actions and emergent behaviors.
The authors build on the framework of  QLM, employing mathematical tools from statistical physics, nonlinear dynamics, and network science.
\begin{itemize}
    \item \textbf{Random Lasers (RLs)}: Systems where light undergoes multiple scattering in disordered media, producing random but coherent lasing modes.
    \item \textbf{Social Random Lasing}: Metaphorical model where social interactions and spontaneous actions resemble light scattering and amplification in RLs.
    \item \textbf{Cognitive Modes}: Analogous to laser modes, representing dominant patterns of opinions or behaviors in a social system.
\end{itemize}

\subsection*{Mathematical Modeling}

\subsubsection*{Lasing Threshold Condition}

The lasing threshold in a random laser corresponds to a minimal level of gain required for self-sustained lasing. Mathematically,
\begin{equation}
\Gamma G > \gamma_c,
\end{equation}
where:
\begin{itemize}
    \item $\Gamma$: Overlap factor between gain and feedback regions.
    \item $G$: Gain coefficient.
    \item $\gamma_c$: Loss coefficient (dissipation).
\end{itemize}

In social terms, this reflects the condition under which spontaneous social actions amplify.

\subsubsection*{Master Equation for Mode Intensities}

The temporal evolution of the intensity $I_n$ of mode $n$ obeys the nonlinear master equation:
\begin{equation}
\frac{dI_n}{dt} = -\gamma_n I_n + G_n I_n - \beta_n I_n^2 + \sum_{m \neq n} \kappa_{nm} I_m I_n,
\end{equation}
where $\gamma_n$: loss for mode $n,$ $G_n$: gain for mode $n,$ $\beta_n$: nonlinear saturation term, $\kappa_{nm}$: mode coupling coefficients.

This equation models competition between social behaviors or narratives, where nonlinear saturation and inter-mode coupling play roles.

\subsubsection*{Applications to Social Phenomena}

\begin{itemize}
    \item \textbf{Collective Actions}: Emergence of dominant modes corresponds to large-scale social movements.
    \item \textbf{Information Overload}: High gains with weak dissipation may model unstable social situations.
    \item \textbf{Narrative Competition}: Coupled nonlinear modes represent competing narratives or ideologies.
\end{itemize}

{\bf Summary:} It seems that random laser physics can effectively simulate social systems characterized by randomness, feedback, and nonlinear dynamics. The mathematical models provide a bridge between optical phenomena and complex social behaviors.

\section{Conventional Versus Random Lasers: Physics and Social Models}

In physics, conventional lasers and \emph{random lasers} differ fundamentally in their operational mechanisms:

\subsubsection*{Physical Perspective}

\begin{itemize}
    \item \textbf{Conventional Lasers:}
    \begin{itemize}
        \item Utilize well-defined optical cavities with mirrors that provide coherent feedback.
        \item The feedback mechanism ensures phase coherence and narrow emission spectra.
        \item Their operation obeys the classical laser threshold condition involving the cavity length and resonance:
        \[
        2L = m \lambda,
        \]
        where \( L \) is the cavity length, \( \lambda \) is the wavelength, and \( m \) is an integer.
        \item High directionality and spatial coherence of emitted light.
    \end{itemize}
    
    \item \textbf{Random Lasers:}
    \begin{itemize}
        \item Lack well-defined mirrors; feedback occurs via multiple scattering in a disordered medium.
        \item Emission arises from a combination of gain and multiple scattering, often producing broader spectra.
        \item Typically less directional and with lower spatial coherence compared to conventional lasers.
        \item Threshold behavior still exists, but it is governed by statistical properties of the disorder.
    \end{itemize}
\end{itemize}

\subsubsection*{Social Perspective: Social Lasers vs. Random Social Lasers}

In the context of \emph{social laser theory}, these physical distinctions inspire different models for social processes:

\begin{itemize}
    \item \textbf{Social Laser (Conventional Laser Analogy):}
    \begin{itemize}
        \item Requires highly organized social structures, such as tightly connected groups or social networks acting as ``resonators''.
        \item Social feedback mechanisms are coherent and strongly reinforcing, akin to optical cavity modes.
        \item Collective actions (``social emissions'') are highly synchronized, often resulting in rapid, mass social movements or revolutions.
        \item Mathematically modeled via resonance conditions:
        \[
        2 L_s = m_s \lambda_s,
        \]
        where \( L_s \) is the effective social network reach, \( \lambda_s \) the dominant narrative wavelength, and \( m_s \) an integer resonance number.
    \end{itemize}
    
    \item \textbf{Random Social Laser (Random Physical Laser Analogy):}
    \begin{itemize}
        \item Arises in disordered, loosely connected social environments where interactions are stochastic.
        \item Social feedback is diffusive and occurs through random social encounters, media fragmentation, or unstructured information flows.
        \item Collective behaviors emerge through the accumulation of random interactions, possibly leading to localized, unpredictable outbursts of social activity.
        \item Analyzed via stochastic models, including percolation theory and random network dynamics, where thresholds depend on connectivity randomness.
    \end{itemize}
\end{itemize}

\subsubsection*{Summary of Differences}

\begin{center}
\begin{tabular}{@{}p{4cm}p{5cm}p{5cm}@{}}
\toprule
 & \textbf{Conventional Laser / Social Laser} & \textbf{Random Laser / Random Social Laser} \\
\midrule
\textbf{Structure} & Highly ordered (cavity or strong social ties) & Disordered, stochastic (scattering or random ties) \\
\textbf{Feedback Mechanism} & Coherent, resonant amplification & Multiple scattering / random interactions \\
\textbf{Threshold Behavior} & Sharp, well-defined threshold & Statistical threshold based on disorder \\
\textbf{Output} & Highly coherent, directional (or synchronized social action) & Diffuse, broad, possibly fragmented social phenomena \\
\textbf{Mathematical Tools} & Resonance models, spectral graph theory & Random matrix theory, percolation, stochastic dynamics \\
\bottomrule
\end{tabular}
\end{center}

{\bf Implications for Social Modeling:}
The random laser model suggests that even in fragmented societies without strong central coordination, collective effects can arise due to stochastic accumulations of weak signals. This insight broadens social laser theory, allowing it to describe decentralized, spontaneous social processes in addition to structured movements.

\section{Color Revolutions Through the Lens of  Social Laser}
\label{SASA}

\subsubsection*{Overview of Color Revolutions}

Color Revolutions are a series of political uprisings in post-Soviet and semi-authoritarian states which was started in the early 2000s, typically named after symbolic colors or objects used by protestors (e.g.~Rose – Georgia 2003; Orange – Ukraine 2004; Tulip – Kyrgyzstan 2005; Velvet – Czech Republic~1989). They share some common traits, as e.g., contested elections, mobilized civic action, use of social media, and rapid spread across countries.

It is important to mention a block of color revolutions known as Arab Spring:  Tunisia – Jasmine Revolution (2010–2011), Egypt – January 25 Revolution (2011), Libya – Uprising Against Gaddafi (2011); led to civil war sparked by protests; Yemen – Yemeni Revolution (2011–2012); Bahrain – Pearl Revolution (2011); Syria – Civil Uprising (2011),
Trigger: Arrest and torture of teenagers for anti-regime graffiti.
escalated into a protracted and brutal civil terminated (hopefully!) in 2024.

Scholars in social and political sciences tried to explained these revolutions through multiple lenses \cite{C1,C2,C3,C4,C5,C6,C7,C8,C9,C10,C11,C12,C13,C14}:
\begin{itemize}
  \item \textbf{Structural and diffusion theories}: point to weak state institutions, authoritarian fragility, and emulation across borders.
  \item \textbf{Civil resistance}: emphasizes that a small active minority (~3.5\%) can topple regimes via nonviolent protest.
  \item \textbf{Media and network mediation}: highlight how traditional and social media and digital tools shaped mobilization and resonance.
\end{itemize}
The views and conclusions of these authors differ very much and no general and consistent theory of color revolutions was constructed. 

\subsubsection*{Modeling of Color Revolutions with Social Laser Theory}

{\bf Empirical Scope:}
Article \cite{laser3} argues that a variety of sociopolitical phenomena, known as color revolutions, display characteristics of 
\emph{Stimulated Amplification of Social Actions}  described by SLT. Key shared traits include: a) rapid, mass collective mobilization lacking centralized leadership or detailed ideology (so color revolutions do not have such leaders as Cromwell, Robespierre, Marx, Lenin, Trotsky, Stalin, Mao Zedong, Fidel Castro and they are not based on advanced ideological systems as marxism or trotskism); b) cascade-like dynamics, marked by exponential growth in participation;
 b)  quick relaxation post-peak, suggesting coherence decay in the social system. 

{\bf Thresholds and Phase Transitions:}
Through a field-thermodynamic lens, Khrennikov \cite{laser5} shows that social systems undergo an abrupt, phase-transition–like move from below-to above-threshold activity once pumping exceeds a critical level. This triggers the cascade behaviour seen in color revolutions.

{\bf Coherence and Relaxation:}
It was pointed out  \cite{laser5}  that, akin to lasers, social systems exhibit high coherence (shared frames, slogans) during stimulated amplification of social actions, followed by a fast decay (quick relaxation) after the surge is exhausted. This mirrors the transient nature 
of many color revolutions.

{\bf Theoretical and Practical Implications:}
This SLT based model of color revolutions offers an operational framework to detect and possibly predict sociopolitical cascades,  stresses the critical role of mass media as both pump and resonator, suggests possible strategies for stabilization - by influencing pump strength or disrupting coherence pathways.

To utilize Khrennikov’s model:
\begin{enumerate}
  \item Identify evidence of agitation (\emph{population inversion}) from media narratives.
  \item Examine coherence (shared slogans, hashtags) for signs of lasing.
  \item Track temporal growth rates to assess threshold crossings.
  \item Measure post-peak relaxation speeds to understand coherence dissipation.
\end{enumerate}
This approach highlights Color Revolutions as manifestations of coherent, stimulated social emission - perfectly consistent with SLT.

\section{Media and Social Mobilization}

In \cite{laser10}  interdisciplinary SLT was applied to analyze media-driven social mobilization, especially amid 
geopolitical conflicts. Rather than literal analogies, the theory is QL: it views individuals as 
social atoms and information units as infons or social photons carrying social energy (emotional or ideological resonance) 
\begin{itemize}
    \item \textbf{SLT in Context of Social Mobilization:}
    \begin{itemize}
\item {\it Social atoms:} individuals share a spectrum of excitability, responding to specific information frequencies.
\item {\it Gain medium:} a group of individuals whose excitability aligns closely (Gaussian distribution) with incoming infons.
\item {\it Social resonators:} echo chambers, such as Facebook groups, that reflect and amplify social energy.
\item {\it Population inversion:} when $>50\%$ of the gain medium receives enough excitation, they begin emitting infons themselves, leading to a cascade of coherent outputs.
\item {\it Spontaneous vs. stimulated emission:} initial reactions may be random, but once coherence is high, responses align with the dominant information vector (cognitive marker).
\item {\it Coherence and echo chambers:} resonators reinforce label-driven reactions over critical analysis, maximizing amplification.
\end{itemize}
\end{itemize}

\subsubsection*{Social Mobilization in Sweden (Feb-May 2022).}

\begin{itemize}
    \item \textbf{Data \& Methods:}
    \begin{itemize}
\item {\it Quantitative media measure:} Content analysis of three major Swedish 
newspapers — {\it Dagens Nyheter, Aftonbladet, and Expressen} - tracking `Ukraine' coverage 
from October 2021 to May 2022. 
\item {\it Qualitative Facebook analysis:} 98 threads from social-media pages of those papers, chosen for $>5$ replies and conflict-focus. Conversation Analysis (CA) and Critical Discourse Analysis (CDA) tracked  two basic cognitive markers, `Ukraine support' - $\alpha,$ challenge and problematization of `Ukraine support' - $\beta.$ 
\end{itemize}
\end{itemize}
 
\begin{itemize}
    \item \textbf{Media Radiation \& Social Mobilization:}
    \begin{itemize}
\item {\it Information tsunami} (March 2022): Newspaper articles mentioning Ukraine tripled between Jan–Feb, 
then peaked at approximately 4,000 in March. This sparked mass emotional responses on Facebook:
High emotional content (emojis, flags, slogans). Posts typically short, with immediate, 
bandwagon-like support; little deliberation. Echo chamber effects: Ukraine-supportive cognitive marker $\alpha$ dominated; critics 
carrying non-supportive cognitive marker $\beta$ were marginalized.
\item {\it Population inversion:} More than half of active participants produced excitatory reactions - likes, shares, emotional posts - triggering social lasing.
\item {\it April–May:} Media radiation lessened (back to Feb levels). Concurrently, dissent grew: increasing cognitive marker $\beta$ - challenges, problematization, and more nuanced exchanges. Facebook threads became more multi-layered and argumentative.
\end{itemize}
\end{itemize}

\begin{itemize}
    \item \textbf{Dynamics of Echo Chambers:}
    \begin{itemize}
\item {\it Coherence maintenance:}
 \begin{itemize}
\item Homogeneous audiences (e.g., DN paid subscribers) produced more predictable and cohesive reactions.
\item Heterogeneous readership (Expressen) showed more challenges, especially among non-Swedish-named participants.
\item Event/person-focused articles yielded more consistent support, unlike debatable content.
\end{itemize}
\item {\it Suppression of dissent:}
\begin{itemize}
\item Early March threads showed rapid attacks on challengers - troll-labelling, doxing threats, censorship by peers or moderators.
\item Spamming (random greetings, off-topic content) emerged as a tactic to reduce coherence and impede dissent.
\item As media pressure fell, dissenters re-emerged; noise (spam) became an effective counter-strategy, at times stronger than reasoned argumentation.
\end{itemize}
\end{itemize}
\end{itemize}

\begin{itemize}
    \item \textbf{Main Findings:}
    \begin{itemize}
\item Social Lasing as Information Warfare: Coordinated, emotionally intense media coverage triggered social lasing. Facebook acted as a social resonator, broadcasting and sustaining the dominant social marker $\alpha.$
\item Temporal alignment: Peaks in media output (March) aligned with spikes in coherent supportive reactions. As media subsided, retreating echo-chambers allowed space for challengers.
\end{itemize}
\end{itemize}

\begin{itemize}
    \item \textbf{Mechanisms of coherence:}
    \begin{itemize}
\item Social resonators and cultural homogeneity boosted resonance.
\item Emotional infons and simplification (´scanning mode', popcorn thinking) drove fast, uniform responses.
\end{itemize}
\end{itemize}

\begin{itemize}
    \item \textbf{Counters to coherence:}
    \begin{itemize}
\item Suppression of dissent through peer or moderator action.
\item Spamming as deliberate disruption.
\item Cognitive $\beta$-marker resurgence tied to reduction in media intensity.
\end{itemize}
\end{itemize}

\begin{itemize}
    \item \textbf{The main output of this case study:}
    \begin{itemize}
\item Demonstrates quantitative-qualitative convergence: systematic media output matches observable patterns in online discourse.
\item Offers a framework for understanding strategic information operations—especially useful in conflict communication contexts.
\item Provides starting points for future work: measuring social energy continuously, testing interventions, 
expanding to other contexts.
\end{itemize}
\end{itemize}

\section{Concluding remarks}

The Social Laser Theory (SLT) offers a unified framework for understanding the emergence of large-scale collective behavior in modern societies by drawing structural analogies to quantum optics and laser physics. Within this framework, the notions of \emph{social atoms}, \emph{social energy}, and \emph{stimulated emission} provide a coherent language for describing how informational and emotional excitations accumulate, synchronize, and ultimately discharge in the form of coordinated social action. By extending the formalism of quantum-like modeling (QLM) to the domain of social fields, SLT bridges micro-level cognitive processes with macro-level sociological phenomena through mechanisms of coherence, resonance, and amplification.

The introduction of the concept of a \emph{social resonator} marks a significant step toward mathematical formalization. Social networks, viewed as resonant cavities for narratives and emotions, can sustain standing informational waves whose stability and coherence depend on the network’s topology and communicative dynamics. The resonance condition
expresses the coupling between the effective social network reach 
and the dominant social wavelength, governed by reinforcement cycles. Spectral graph theory, graph wavelets, and topological data analysis (TDA) provide rigorous pathways to quantify and simulate these phenomena. These methods not only enable identification of critical frequencies at which social narratives become self-reinforcing but also open the door to predictive diagnostics of social coherence and potential mobilization thresholds.

In this light, SLT establishes an interface between the structural properties of communication networks and the dynamic propagation of meaning. The emerging picture is that social systems, much like optical systems, can transition from incoherent fluctuation to coherent emission once resonance conditions are satisfied. Such coherence manifests in synchronized collective behavior—mass protests, ideological cascades, or viral movements—representing a phase-like transition in the socio-informational field.

A particularly relevant dimension of SLT in contemporary society concerns the phenomenon of \emph{information overload}. The digital environment, characterized by incessant flows of heterogeneous data, produces conditions reminiscent of energetic pumping in physical lasers. Although overload is typically viewed as detrimental—leading to cognitive fatigue, decision paralysis, and social fragmentation—SLT reframes it as a double-edged process. On one hand, the excess of incoherent information disperses individual attention and undermines rational judgment; on the other, it establishes the energetic background necessary for selective resonance. When certain narratives are emotionally aligned, repetitive, and phase-synchronized across media channels, they stand out against the informational noise, facilitating the coherent excitation of the social field.

In this sense, information overload becomes both a symptom of the digital age and a precondition for quantum-like social transitions. It provides the ``pumping energy’’ that can drive populations toward states of heightened excitation, preparing the ground for stimulated emission of social energy—manifested as coordinated collective action. SLT thus elucidates how informational excess, far from merely overwhelming societies, may actively contribute to their capacity for rapid, synchronized transformation.

From a broader perspective, these insights carry implications for media ecology, political communication, and social governance. Understanding the resonance conditions and coherence thresholds within social networks could enable the development of diagnostic tools for detecting approaching critical states of social excitation—offering both opportunities for constructive mobilization and warnings against destabilizing cascades.

Future research should aim at refining the mathematical apparatus of SLT, integrating empirical data from network dynamics, sentiment analysis, and media propagation models. The coupling of quantum-like probabilistic frameworks with data-driven modeling could provide operational ways to measure the effective reach 
dentify dominant narrative frequencies $\lambda_s$
and estimate the parameters governing social coherence.

Ultimately, SLT invites us to reconsider modern society as an open, self-organizing resonant system where information, emotion, and meaning interact according to principles reminiscent of quantum coherence. In such a system, understanding and managing resonance phenomena may become as essential to social stability as energy regulation is in physical systems.

\section*{Acknowledgments} 
This research was partially supported by  the EU-grant CA21169 (DYNALIFE), by  JST, CREST Grant Number JPMJCR23P4, Japan, and visiting professor fellowship at Ritsumeikan University (April 2025). The author would like to thank prof. Miho Fuyama for fruitful discussions and hospitality.

\section{Complementary Material}
\label{CM}

\subsection{Social Atom and Social Molecule}
\label{SA}

The concepts of the social atom and social molecule emerged in the 19th century as part of the broader effort to apply scientific and particularly physicalist analogies to the understanding of society. Inspired by the success of atomic theory in chemistry and physics, early sociologists and social philosophers began to conceptualize individuals and their groupings in terms of atoms and molecules, aiming to bring the precision and rigor of the natural sciences to the study of social structures.

{\bf  Auguste Comte and Early Positivism.}
One of the first to draw analogies between physical and social entities was Auguste Comte, the founder of positivism. Although he did not use the exact term social atom, Comte advocated for a science of society grounded in empirical laws, similar to physics and chemistry. His work laid the conceptual groundwork for viewing individuals as basic units of social analysis.

{\bf  Gabriel Tarde} (Late 19th Century).
French sociologist Gabriel Tarde \cite{Tarde1,Tarde2} explicitly used atomistic analogies in his work. He viewed individuals as social atoms whose interactions produced complex social structures, much as atoms form molecules and compounds. Tarde emphasized imitation and innovation as forces that bind social atoms into larger units, anticipating ideas of information transfer and social contagion.

{\bf Emile Durkheim and Holism.}
In contrast to Tarde's atomism, Emile Durkheim \cite{Durkheim} criticized the reduction of society to individual elements. While he acknowledged individuals as the basic constituents of society, he argued that social facts were ``sui generis'' and could not be reduced to individual psychology. Durkheim's view parallels the shift from atomistic to holistic models in physics (e.g., field theories), and his emphasis on collective consciousness can be seen as an early form of ``social molecule'' thinking, where the whole is more than the sum of its parts.

{\bf Sociophysics and the 20th Century.}
The analogy between atoms and individuals gained new traction in the 20th century, especially within sociophysics and mathematical sociology. Scholars  such as L`eon Winiarski 
, Louis Bachelier, and later Ettore Majorana, and Serge Galam treated individuals as particles obeying statistical laws, modeling crowd behavior, opinion dynamics, and social transitions with methods from statistical mechanics (see, e.g., 
\cite{Winiarski1,Winiarski2,Majorana1,Majorana2}).

{\bf Jacob L. Moreno and Sociometry} (1930s–1950s), \cite{Moreno1,Moreno2,Moreno3,Moreno4,Moreno5,Moreno6,Moreno7} 
A major turning point came with Jacob Moreno, who introduced the term social atom in a psychological and sociometric context. For Moreno, the social atom referred to the smallest unit of human social structure: the individual along with their immediate social relationships. He used this concept in therapeutic settings, emphasizing the configuration of significant interpersonal bonds, rather than the isolated individual.

{\bf Quantum and Information-Theoretic Extensions} (21st Century)
In recent decades, especially in SLT, the notion of the social atom has been revitalized through QLM. In this new framework, the social atom is treated as a QL system with discrete internal energy states, capable of probabilistic transitions and entanglement with other social atoms. Groups of such atoms form social molecules through coherence and interaction, capable of collective behavior under the influence of quantized information signals—marking the conceptual shift from metaphor to formal analogy.

From its origins in philosophical metaphor to its modern formalization in mathematical and QLMs, the idea of the social atom has remained a powerful tool for understanding the individual's place in complex social systems. The notion of the social molecule, while less frequently invoked, complements this view by offering a framework for emergent collective structures, paving the way for theories like the Social Laser, where coherence and amplification arise from quantum-inspired social dynamics.

\subsection{The Evolution of the Notions of Psychic, Mental, and Social Energy}
\label{energy}

The notions of psychic, mental, and social energy emerged at the intersection of psychology, philosophy, and sociology, developing over centuries as metaphoric frameworks for understanding the invisible forces driving human behavior, thought, and collective action. Though none of these terms describe physical energies in the strict scientific sense, their metaphorical utility has endured across disciplines, shaping models of individual and social dynamics.

\subsubsection*{Psychic and Mental Energy in Early Psychology}
The concept of psychic energy can be traced to 19th-century psychology, most notably to Sigmund Freud's psychoanalytic theory. Freud introduced the idea of libido - a form of psychic energy — within his broader theory of the mind. In Three Essays on the Theory of Sexuality \cite{Freud_sex1,Freud_sex2} and Beyond the Pleasure Principle \cite{Freud_pleasure1,Freud_pleasure2}, Freud described libido as the dynamic, drive-based force behind human motivation and behavior. He envisioned the mind as an energy economy, where repressed desires and instincts created tensions requiring release or transformation.

Freud’s conceptualization drew from earlier metaphors in physics, such as Helmholtz's principle of conservation of energy, applied metaphorically to mental processes. The Freudian model was subsequently elaborated upon by Carl Jung, who retained the notion of psychic energy but broadened it beyond the sexual \cite{Jung}. For Jung, psychic energy (or libido) encompassed a wider range of motivational forces, including the drive for individuation—the process of self-realization.

Concurrently, William James had articulated similar ideas without using the energy metaphor explicitly. In The Principles of Psychology \cite{James}, James discussed the mind's limited resources, effort, and volition, laying groundwork for later cognitive interpretations of mental energy.

\subsubsection*{From Psychology to Sociology: The Birth of Social Energy}
While psychic and mental energy primarily described individual psychological dynamics, the term social energy emerged in sociological and political thought. Emile Durkheim, though not using the phrase directly, offered foundational insights into collective consciousness and the role of rituals in generating emotional resonance across groups. In \cite{Durkheim} Durkheim  suggested that collective effervescence—intense shared emotion—was a key force in societal cohesion, arguably an early form of the 
``social energy'' concept.

In the 20th century, Norbert Elias \cite{Elias} emphasized the long-term transformation of emotional regulation in social groups. Though not couched in terms of energy, Elias's notion of controlled affect and societal norms implicitly ties into the energy metaphor.

The term social energy itself gained more currency in cultural theory and critical sociology. Jean Baudrillard \cite{Baudrillard} hinted at the symbolic value and energy of social signs and consumer behaviors. More explicitly, the philosopher and media theorist Boris Groys \cite{Groys} discussed cultural energy in relation to aesthetics and politics, emphasizing how collective attention and emotion serve as reservoirs of power in modern societies.

\subsubsection*{Contemporary Uses and Interdisciplinary Expansions}
Recent theoretical models have sought to formalize these metaphoric energies. In cognitive science, Roy Baumeister's ego depletion theory described self-control and decision-making as relying on a finite resource akin to mental energy (see Baumeister et al. \cite{Baumeister}). Though the metaphor remains debated empirically, it has influenced research in behavioral economics and decision theory.

In SLT framework, social energy is treated as a quantifiable analogue to physical energy, operating in a informational space and driving coherent collective behavior under certain conditions.

In popular and therapeutic discourses, the terms psychic and mental energy remain widespread, albeit often detached from their theoretical origins. They serve as intuitive metaphors in wellness culture, motivational psychology, and even in political rhetoric where collective momentum or energy describes the intensity of mass participation.

{\bf Summary:} The trajectory of psychic, mental, and social energy reflects a persistent attempt to make sense of invisible yet powerful dynamics—whether intrapsychic conflicts, cognitive exertion, or collective passions. Though not rigorously defined in scientific terms, these metaphors have deeply influenced how both scholars and laypeople conceptualize human motivation and social behavior. Contemporary efforts to model these energies—ranging from cognitive psychology to QL theories—continue to wrestle with their metaphorical power and scientific viability.

\subsection{Hermann Haken: Social Science and Relation to Social Laser Theory}
\label{Haken}

Hermann Haken, the founder of \emph{synergetics}, developed a rigorous mathematical framework for analyzing self-organization in complex systems far from equilibrium. Though initially focused on physical systems such as lasers, Haken explicitly extended these ideas to social and economic systems (see \cite{Haken1983,Haken1997,Haken2006}).

\subsubsection*{Synergetics and Social Systems}

In his seminal book \emph{Synergetics: An Introduction} \cite{Haken1983}, Haken dedicated entire sections to the application of synergetics in social contexts. Key concepts include:
\begin{itemize}
    \item \textbf{Order Parameters:} Macroscopic variables governing system behavior, e.g., collective opinions.
    \item \textbf{Slaving Principle:} Microscopic variables (e.g., individual actions) are ``enslaved'' by these order parameters.
    \item \textbf{Instability and Phase Transitions:} Sudden social changes, such as revolutions or opinion shifts, may correspond to critical points.
\end{itemize}

\textbf{Selected citation} from \cite[p.~275]{Haken1983}:
\begin{quote}
``It is particularly interesting to apply the concept of order parameters to social systems, since there, macroscopic variables such as the average attitude of a population may enslave the dynamics of individual actions.''
\end{quote}

\subsubsection*{Laser Metaphors in Haken's Works}

In his discussion of lasers, Haken often draws analogies to social processes:
\begin{itemize}
    \item \textbf{Threshold effects:} Systems require external driving beyond a critical threshold to exhibit coherent behavior.
    \item \textbf{Amplification and Coherence:} Weak perturbations may get amplified through collective interactions.
\end{itemize}

For instance, in \cite[p.~307]{Haken1983}, Haken notes:
\begin{quote}
``The laser can serve as a prototype for understanding the emergence of coherence in many-particle systems, including models of society.''
\end{quote}

\subsubsection*{Did Haken Propose Social Laser Theory?}

Haken did \emph{not} introduce the term \emph{social laser} or formulate a detailed model based on laser physics for social systems. Rather, he employed \emph{laser-like metaphors} to explain general mechanisms of self-organization and coherence.

\textbf{Key Distinction:}
\begin{itemize}
    \item Haken's approach focused on \emph{general self-organization}, without mapping specific laser components (resonators, pumping) to social processes.
    \item Modern \emph{social laser theory}, developed later (notably by A. Khrennikov and collaborators), explicitly formalizes this mapping, describing social networks as resonators and media as pumping sources.
\end{itemize}

{\bf Summanry:} Haken’s pioneering work laid a mathematical foundation for studying coherence in social systems via synergetics. However, the explicit \emph{social laser} concept emerged only later, building on his metaphors but developing them into a detailed, formalized theory.

\subsection{Social Laser Versus Maser}
\label{Tdis}

We start with a terminological remark. In physics, one distinguish lasers and masers. In fact, these are devices based 
on the same principles and different only the wavelengths of emitted radiation. Laser is the abbreviation of 
Light Amplification by Stimulation Emission of Radiation, while Maser is the abbreviation of  
Microwave Amplification by Stimulation Emission of Radiation. Hence, laser emits radiation with shorter wavelengths 
(or higher frequency) than maser, or in terms of energy of photons, laser emits photons of higher energy than maser.  

In our social laser framework the term `laser' was invented to highlight the analogy with physical laser. There is no natural definition of social frequency and social wavelength, the whole theory is structured in terms of social energy. 

\subsection{Social Planck Constant? Social Frequency and Wavelength?}
\label{h}

In physics, energy and frequency are connected by the Einstein formula: 
the Einstein formula, 
\begin{equation}
\label{EIN}
\nu= E/h, 
\end{equation}
where $h$ is the Planck constant; 
the wave length is given by 
\begin{equation}
\label{EINa}
\lambda= hc/E. 
\end{equation}
One might proceed in this way for social processes and starting with operationally determined 
social energy, define social frequency  and  wavelength. However, such approach 
would be fruitful if we would be able to determine social (mental) analogs of the Planck constant and 
light velocity. If fact, the maximal velocity of  media message can be taken equal to light 
velocity. So, the main problem is determination of the minimal and indivisible quantum of social action -
social (mental) Planck constant. This is a complex problem. 

Therefore, we prefer to proceed solely with social energy and, instead of distinguishing ``social lasers and  masers'',
to speak about social lasers of high and low energy.

\subsection{Maser Stars in the Cosmos} 
\label{cosmic} 

In the vast reaches of interstellar space, long before humans developed microwave amplification in laboratories, the universe had already mastered the art. Certain stars, particularly those in late stages of stellar evolution, emit intense, naturally occurring maser radiation—microwave amplification by stimulated emission of radiation. These maser stars serve as some of the most fascinating and useful cosmic laboratories in the sky.

Maser stars are typically evolved, cool, late-type stars, such as Mira variables, semi-regular variables, and red supergiants, that exhibit strong maser emission in their extended atmospheres and circumstellar envelopes. As these stars shed mass through slow, dense stellar winds, they form vast clouds of molecules—especially water vapor ($H_2O$), hydroxyl ($OH$), and silicon monoxide ($SiO$)—that can emit intense, coherent microwave radiation under the right physical conditions.

{\it The maser process in these environments arises from a combination of the following conditions:}

Population inversion, where more molecules are in an excited energy state than in a lower one (due to radiative pumping or collisions),

Long path lengths in the circumstellar material, allowing significant amplification,

Velocity coherence, where the relative motion of gas enhances the probability of stimulated emission along specific lines of sight.

These conditions allow maser emissions to become extremely bright, often outshining the entire host star in the maser wavelength bands, even though they originate from relatively small volumes of space.

{\bf Famous Maser Stars.} A number of well-known stars exhibit strong maser activity. Here are some prominent examples:

Mira (o Ceti): A pulsating red giant and prototype of Mira variables. Mira displays both $OH$ and $SiO$ masers, which vary with the star's pulsation cycle.

W Hydrae: A semi-regular variable star that shows strong water ($H_2O$) and silicon monoxide ($SiO$) maser emissions, providing insight into stellar pulsations and mass loss.

VX Sagittarii: A highly luminous red supergiant, known for its intense $OH, SiO,$ and $H_2O$ masers. It is one of the brightest stellar maser sources in the sky and a key object for VLBI studies.

IRC+10216 (CW Leonis): A carbon-rich AGB star surrounded by a thick envelope of dust and molecules. It is one of the brightest infrared sources in the sky and exhibits SiO masers among others.

\medskip

{\bf Summary:} Maser stars represent one of nature's most remarkable coincidences: a stellar environment perfectly tuned to amplify microwaves across light-years of space. Observing them is like eavesdropping on the inner workings of stars through naturally occurring cosmic beacons - quiet, persistent voices whispering through the dark of the galaxy

Portions of this manuscript were drafted or edited with the assistance of ChatGPT, a large language model developed by OpenAI.
This portions are based on the articles of the author and his coauthors.
 The author is solely responsible for the final content.


\begin{thebibliography}{199}
\bibitem{Aerts1}
D. Aerts and J. Broekaert,
``Quantum Cognition: A Quantum-Theoretic Model of Cognitive Contextuality,''
\emph{Journal of Mathematical Psychology}, 39, no. 3, pp. 346–374, 1995.

\bibitem{Aerts2}
D. Aerts, J. Broekaert, and S. Smets,
``The Liar-Paradox in a Quantum Mechanical Perspective,''
\emph{Foundations of Science},  4, no. 2, pp. 115–132, 1999.

\bibitem{Aerts3}
D. Aerts, J. Broekaert, and S. Smets,
``Quantum Structure in Cognition: A Theoretical Framework,''
\emph{Foundations of Science}, 4, no. 2, pp. 133–157, 1999.

\bibitem{Aerts4} Aerts, D., Broekaert, J., Gabora, L., \& Sozzo, S. (2013). Quantum structure and human thought. Behavioral and Brain Sciences, 36(3), 274-276.
\bibitem{Aerts} Aerts, D.; Gabora, L.; Sozzo, S. Concepts and their dynamics: A quantum-theoretic modeling of human thought. Top. Cogn. Sci. 2013, 5, 737-772

\bibitem{laser12} Alodjants, A. P., Bazhenov, A. Y., Khrennikov, A. Y., \&
Bukhanovsky, A. V. (2022). Mean-field theory of social laser. \textit{%
Scientific Reports}, 12(1), 8566.

\bibitem{laser13} Alodjants, A., Zacharenko, P., Tsarev, D., Avdyushina, A., Nikitina, M., Khrennikov, A., \& Boukhanovsky, A. (2023). Random lasers as social processes simulators. Entropy, 25(12), 1601.

\bibitem{QBIOP} Arndt, M., Juffmann, T., and Vedral, V.: Quantum physics meets biology. HFSP J \textbf{3}  386--400 {2009}

\bibitem{Asano} Asano M, Basieva I, Khrennikov A, Ohya M, Tanaka Y, Yamato I. A model of epigenetic evolution based on theory of open quantum systems. Syst Synth Biol. 2013 Dec;7(4):161-73.
\bibitem{QIB} M. Asano, I. Basieva, A. Khrennikov, M. Ohya, Y. Tanaka, I. Yamato Quantum Information Biology: from information interpretation of quantum mechanics to applications in molecular biology and cognitive psychology. {\it Found. Phys.} 
45, N 10, 1362-1378 (2015).
\bibitem{QL3} Asano, M., Khrennikov, A., Ohya, M., Tanaka, Y. and Yamato, I.: Quantum Adaptivity in Biology: from Genetics to Cognition. Springer, Heidelberg-Berlin-New York (2015).

\bibitem{Bagarello1} Bagarello, F., Basieva, I., Pothos, E. M., \& Khrennikov, A. (2018). Quantum like modeling of decision making: Quantifying uncertainty with the aid of Heisenberg–Robertson inequality. Journal of mathematical psychology, 84, 49-56.
\bibitem{Bagarell2} Bagarello, F. (2019). \textit{Quantum Concepts in the
Social, Ecological and Biological Sciences}. Cambridge University Press:
Cambridge, UK.
\bibitem{Bagarello3} Bagarello, F., Gargano, F., \& Oliveri, F. (2023). Quantum tools for macroscopic systems. Cham, Switzerland: Springer.
\bibitem{Bagarello4} Bagarello, F., Gargano, F., Gorgone, M., \& Oliveri, F. (2023). Spreading of information on a network: a quantum view. Entropy, 25(10), 1438.


\bibitem{Barahona} Barahona, M., Pecora, L.M. (2002). Synchronization in small-world systems. Phys. Rev. Lett. 89(5), 054101.

\bibitem{Behti} Basieva, I., Cervantes, V. H., Dzhafarov, E. N., \& Khrennikov, A. (2019). True contextuality beats direct influences in human decision making. Journal of Experimental Psychology: General, 148(11), 1925.
\bibitem{BioBas} I. Basieva, A. Khrennikov, M. Ozawa, Quantum-like modeling in biology with open quantum systems and instruments,
Biosystems, 201, 2021, 104328.

\bibitem{Baumeister} Baumeister, R. F., Bratslavsky, E., Muraven, M., \& Tice, D. M. (1998). Ego depletion: Is the active self a limited resource? Journal of Personality and Social Psychology, 74(5), 1252–1265. https://doi.org/10.1037/0022-3514.74.5.1252

\bibitem{Baudrillard} Baudrillard, J. (1975). The Mirror of Production (M. Poster, Trans.). St. Louis, MO: Telos Press.

\bibitem{C1}
Beissinger, M.R. (2007). Structure and Example in Modular Political Phenomena: The Diffusion of Bulldozer/Rose/Orange/Tulip Revolutions. \textit{Perspectives on Politics}, 5(2), 259–276.

\bibitem{Boole1}  Boole, G.:  On the theory of probabilities. Phil. Trans. Royal Soc. \textbf{152}, 225--242  (1862)
\bibitem{Boole2}  Boole, G.:   An Investigation of the Laws of Thought. New York, Dover (1958)

\bibitem{C2}
Bunce, V.J. \& Wolchik, S.L. (2006). International diffusion and postcommunist electoral revolutions. \textit{Communist and Post-Communist Studies}, 39(3), 283–304.

\bibitem{C3}
Bunce, V.J. \& Wolchik, S.L. (2006). Favorable Conditions and Electoral Revolutions. \textit{Journal of Democracy}, 17(4), 5–18.

\bibitem{Bruza}  Bruza, P. D., Kitto, K., Ramm, B. J., and Sitbon, L. (2015). A probabilistic framework for analysing the compositionality of conceptual combinations. J. Math. Psych., 67, 26-38.

\bibitem{Bruza1} Bruza, P. D., Wang, Z., \& Busemeyer, J. R. (2015). Quantum cognition: a new theoretical approach to psychology. Trends in cognitive sciences, 19(7), 383-393.

\bibitem{Bruza2} Bruza, P. D., Fell, L., Hoyte, P., Dehdashti, S., Obeid, A., Gibson, A., \& Moreira, C. (2023). Contextuality and context-sensitivity in probabilistic models of cognition. Cognitive Psychology, 140, 101529.

\bibitem{Busemeyer1} Busemeyer, J.R.; Bruza, P.D. Quantum Models of Cognition and Decision; Cambridge University Press: Cambridge, 
UK, 2012; 2nd Edition, 2024.
\bibitem{Busemeyer2} Busemeyer, J. R., \& Pothos, E. M. (2013). Theoretical and Empirical Aspects of Quantum Models of Cognition. Topics in Cognitive Science, 5(4), 672–688.
\bibitem{Busemeyer3} Busemeyer, J. R., \& Wang, Z. (2014). Quantum Cognition: Key Issues and Discussion. Topics in Cognitive Science, 6(1), 43–46.
\bibitem{Busemeyer4} Busemeyer, J. R., Wang, Z., Khrennikov, A., \& Basieva, I. (2014).
Applying quantum principles to psychology. \textit{Physica Scripta}, \textit{T163}, 014007.
\bibitem{Busemeyer5} Busemeyer, J. R. (2023). Measurement Models in Quantum Cognition. In The Quantum-Like Revolution: A Festschrift for Andrei Khrennikov (pp. 269-279). Cham: Springer International Publishing.

\bibitem{C4}
Chenoweth, E. \& Stephan, M. (2011). \textit{Why Civil Resistance Works}. Columbia University Press.


\bibitem{V4} Chiolerio, A., Vitiello, G., Dehshibi, M. M., \& Adamatzky, A. (2023). Living plants ecosystem sensing: A quantum bridge between thermodynamics and bioelectricity. Biomimetics, 8(1), 122.


\bibitem{Chung} Chung, F. (1997) \emph{Spectral Graph Theory}. American Mathematical Society, Providence, RI.

\bibitem{C5}
D' Anieri, P. (2005). What has changed in Ukrainian politics? Assessing the implications of the Orange Revolution. \textit{Problems of Post-Communism}, 52(5), 82–91.

\bibitem{Durkheim} Durkheim, E. (1995). The Elementary Forms of Religious Life (K. E. Fields, Trans.). New York, NY: The Free Press. (Original work published 1912)

\bibitem{Elias} Elias, N. (1994). The Civilizing Process (E. Jephcott, Trans.). Oxford, UK: Blackwell Publishing. (Original work published 1939)


\bibitem{EPR} Einstein, A.; Podolsky, B.; Rosen, N.   Can quantum-mechanical description of physical reality be considered complete? \emph{Phys. Rev.} {\bf 1935}, {\it 47}, 777--780.

\bibitem{C6}
Fairbanks Jr, C.H. (2004). Georgia’s Rose Revolution. \textit{Journal of Democracy}, 15(2), 110–124.

\bibitem{Freud_sex1}  Freud, S. Drei Abhandlungen zur Sexualtheorie. Leipzig \& Vienna: Fran Deuticke, 1905. 
\bibitem{Freud_pleasure1} Freud, S. Jenseits des Lustprinzips. Leipzig, Vienna \& Zurich: Internationaler 
Psychoanalytischer Verlag, 1920. 
\bibitem{Freud_sex2} Freud, S. (1905/1953). Three Essays on the Theory of Sexuality.  In J. Strachey (Ed.), The Standard Edition of the Complete Psychological Works of Sigmund Freud (Vol. VII, pp.123-243). London: Hogarth Press \& The Institute of Psycho-Analysis, 1953. 
\bibitem{Freud_pleasure2} Freud, S. (1920/1955). Beyond the Pleasure Principle.  In J. Strachey (Ed.), The Standard Edition of the Complete Psychological Works of Sigmund Freud (Vol. XVIII, pp. 1-64). London: Hogarth Press \& The Institute of Psycho-Analysis, 1955.

\bibitem{Fuyama} M. Fuyama, A. Khrennikov, M. Ozawa (2025) Quantum-like cognition and decision making in the light of quantum measurement theory, accepted to Philosophical Transactions A, http://arxiv.org/abs/2503.05859 .

\bibitem{Groys} Groys, B. (2008). Art Power. Cambridge, MA: MIT Press.


\bibitem{Gunji1} Gunji YP, Sonoda K, Basios V. Quantum cognition based on an ambiguous representation derived from a rough set approximation. Biosystems. 2016 Mar;141:55-66. 
\bibitem{Gunji2}  Gunji YP, Shinohara S., Haruna T., Basios V. Inverse Bayesian inference as a key of consciousness featuring a macroscopic quantum logical structure. Biosystems. 2017,  ;152:44-65.
\bibitem{Gunji3} Gunji, Y.-P., \& Haruna, T. (2022). Concept Formation and Quantum-like Probability from Nonlocality in Cognition, Cognitive Computation. 14, 1328-1349.
\bibitem{Gunji4} Gunji, Y.-P., \& Nakamura, K. (2022). Psychological Origin of Quantum Logic: An Orthomodular Lattice Derived from Natural-Born Intelligence without Hilbert Space. BioSystems 215-216, 104649.


\bibitem{James} James, W. (1890). The Principles of Psychology (Vols. 1–2). New York, NY: Henry Holt and Company.

\bibitem{Jung} Jung, C. G. (1928). On Psychic Energy. In Contributions to Analytical Psychology (H. G. Baynes, Trans., pp. 3-66). London, UK: Kegan Paul, Trench, Trubner \& Co.
\bibitem{Haken1983} Haken, H. \emph{Synergetics: An Introduction. Nonequilibrium Phase Transitions and Self-Organization in Physics, Chemistry and Biology}. 3rd ed., Springer-Verlag, Berlin, Heidelberg, 1983.
\bibitem{Haken1997} Haken, H. \emph{Principles of Brain Functioning: A Synergetic Approach to Brain Activity, Behavior and Cognition}. 2nd ed., Springer, Berlin, Heidelberg, 1997.
\bibitem{Haken2006} Haken, H. \emph{Information and Self-Organization: A Macroscopic Approach to Complex Systems}. 3rd ed., Springer, Berlin, Heidelberg, 2006.


\bibitem{H} Hameroff, S. (1994). Quantum coherence in microtubules. A neural
basis for emergent consciousness? \textit{Journal of Consciousness Studies}%
\emph{,} 1, 91--118.


\bibitem{HavenB}  Haven, E. (2005).  Pilot-wave theory and financial option pricing.
\textit{Int. J. Theor. Phys.} 44 (11), 1957--1962.
\bibitem{QPOL1a} Haven, E.  (2019). Quantum mechanical pragmatic rules and social science. \textit{Activitas Nervosa Superior}
61(1), 83--85.

\bibitem{Haven2009} Haven, E. and Khrennikov, A. (2009). Quantum mechanics and violation of the sure-thing principle: the use of probability interference and other concepts. \textit{J. Math. Psych.} 53, 378--388.
\bibitem{Haven} Haven, E.; Khrennikov, A. (2013). \textit{Quantum Social
Science}. Cambridge University Press: Cambridge, UK.
\bibitem{HAVKHRQ}  Haven, E. and Khrennikov, A. (2016).  Quantum probability and the mathematical modelling of decision-making. \textit{Phil. Trans. Royal Soc. A} 374(2058), 20150105.
\bibitem{HAVKHRQ1} Haven, E., and Khrennikov, A. (2016).  Statistical and subjective interpretations of probability in 
QLMs of cognition and decision making.  \textit{J. Math. Psych.} 74, 82--91.
\bibitem{QLH} Haven, E.,  Khrennikov, A. and Robinson, T. R.   (2017). Quantum Methods in Social Science: A First Course. WSP, Singapore.
\bibitem{handbook}  Haven, E. and Khrennikov, A.  (2017). The Palgrave handbook of quantum models in social science. Macmillan Publishers Ltd:  London. 
\bibitem{Polina2} Haven, E. \& Khrennikova, P. (2018). A
quantum-probabilistic paradigm : non-consequential reasoning and state
dependance in investment choice. \textit{Journal of Mathematical Economics},
78, 186-197. 

\bibitem{Igamberdiev1} Igamberdiev, A. U. (1993). Quantum mechanical properties of biosystems: a framework for complexity, structural stability, and transformations. Biosystems, 31(1), 65-73.
\bibitem{Igamberdiev2} Igamberdiev A. U. (2004),  Quantum computation, non-demolition measurements, and reflective control in living systems. Biosystems 77, 47-56.
\bibitem{Igamberdiev1b} Igamberdiev, A. U., \& Shklovskiy-Kordi, N. E. (2017). The quantum basis of spatiotemporality in perception and consciousness. Progress in biophysics and molecular biology, 130, 15-25.
\bibitem{Igamberdiev2a} A. U. Igamberdiev, J. E. Brenner, Mathematics in biological reality: The emergence of natural computation in living systems Biosystems 204, 2021, 104395.



\bibitem{C7}
Ishigooka, K. (2006). ‘Democratic Revolution in Colors’: Ukraine through the Prism the 2006 Parliamentary Election. \textit{Annals of the Japanese Association for Russian and East European Studies}, 35, 46–59. doi:10.5823/jarees.2006.46

\bibitem{Haven11} Ishio, H., Haven. E. (2009). Information in asset pricing. 
\textit{Annalen der Physik}, 521(1), 33-44.

\bibitem{kahneman1979} Kahneman, D., \& Tversky, A. (1979). Prospect theory: An analysis of decision under risk.
\textit{Econometrica}, 47(2), 263--291.

\bibitem{kahneman1986}
Kahneman, D., \& Tversky, A. (1986).
Rational choice and the framing of decisions.
\textit{Journal of Business}, 59(4), S251--S278.


\bibitem{C8}
Kalandadze, K. \& Orenstein, M.A. (2009). Electoral Protests and Democratization Beyond the Color Revolutions. \textit{Comparative Political Studies}, 42(9), 1179–1205.


\bibitem{KHC1} Khrennikov, A. (1999). Classical and quantum mechanics on
information spaces with applications to cognitive, psychological, social and
anomalous phenomena. \textit{Foundations of Physics}, 29, 1065--1098.

\bibitem{INT} Khrennikov, A.~Yu. (1999)(2nd edition: 2009) \textit{%
Interpretations of Probability}. VSP Int. Sc. Publishers: Utrecht/Tokyo
(1999); De Gruyter: Berlin (2009).

\bibitem{QL0} Khrennikov, A.(2004). \textit{Information Dynamics in
Cognitive, Psychological, Social, and Anomalous Phenomena}. Series:
Fundamental Theories of Physics. Kluwer, Dordrecht.

\bibitem{UB_KHR} Khrennikov, A. (2010). \textit{Ubiquitous Quantum
Structure: From Psychology to Finances}. Springer: Berlin/Heidelberg,
Germany; New York, NY, USA.

\bibitem{AndreiAuman} Khrennikov, A. (2015). Quantum version of Aumann's
approach to common knowledge: sufficient conditions of impossibility to
agree on disagree. \textit{Journal of Mathematical Economics}, 60, 89-104.

\bibitem{PLOS} Khrennikov, A., Basieva, I., Dzhafarov, E.N., and Busemeyer,
J. R. (2014). Quantum models for psychological measurements: An unsolved
problem. \textit{PLOS One} 9, Art. e110909.

\bibitem{laser1} Khrennikov, A. (2015). Towards information lasers. \textit{%
Entropy}, 17(10), 6969-6994.
\bibitem{laser2} Khrennikov, A. (2016). `Social Laser': action amplification
by stimulated emission of social energy. \textit{Philosophical Transactions
of the Royal Society A: Mathematical, Physical and Engineering Sciences},
374(2058), 20150094.
\bibitem{laser3} Khrennikov, A. (2018). Social laser model: from color
revolutions to Brexit and election of Donald Trump. \textit{Kybernetes},
47(2), 273-288.
\bibitem{laser4} Khrennikov, A., Alodjants, A., Trofimova, A., \& Tsarev, D.
(2018). On interpretational questions for quantum-Like modeling of social
lasing. \textit{Entropy}, 20(12), 921.
\bibitem{laser6} Khrennikov, A., Toffano, Z., \& Dubois, F. (2019). Concept
of information laser: from quantum theory to behavioural dynamics. \textit{%
The European Physical Journal Special Topics}, 227, 2133-2153.
\bibitem{laser5} Khrennikov, A. (2020). \textit{Social laser: application of
quantum information and field theories to modeling of social processes.}
Jenny Stanford Publishing, Singapore. 
\bibitem{laser5a} Khrennikov, A. (2020). Social laser theory: 
From quantum physics to collective human behavior. Entropy, 22(11), 1229. 
\bibitem{laser7} Khrennikov, A. (2020). Social laser model for the bandwagon
effect: Generation of coherent information waves. \textit{Entropy}, 22(5),
559.
\bibitem{laser8} Khrennikov, A., Haven, E. (2020). Quantum-like modeling:
From economics to social laser. \textit{Asian Journal of Economics and
Banking}, 4(1), 87-99.
\bibitem{covid}  Khrennikov, A. (2021). Ultrametric diffusion equation on energy landscape to model disease spread in hierarchic socially clustered population. Physica A: Statistical Mechanics and its Applications, 583, 126284.


\bibitem{NL} A. Khrennikov, A.(2019). Get rid of nonlocality from quantum physics. Entropy, 21(8), 806.
\bibitem{NLa} Khrennikov, A. (2020). Two faced Janus of quantum nonlocality. Entropy  \textbf{22}, 303 (2020)

\bibitem{laser9} Khrennikov, A. (2023). Coherent decision making stimulated
within the social laser: open quantum systems framework. \textit{%
Philosophical Transactions of the Royal Society A}, 381(2252), 20220294.

\bibitem{Khrennikov} Khrennikov A. (2023). Open Systems, Quantum
Probability, and Logic for Quantum-like Modeling in Biology, Cognition, and
Decision-Making. \textit{Entropy}. 25(6):886.

\bibitem{Open_QLM} Khrennikov, A. (2023). \textit{Open Quantum Systems in
Biology, Cognitive and Social Sciences}, Springer.

\bibitem{Polina1} Khrennikova, P. (2025). Measuring contextuality in
investment preferences. \textit{Annals of Operations Research}
https://doi.org/10.1007/s10479-025-06630-8. 

\bibitem{K} A. N. Kolmogoroff (1933). \textit{Grundbegriffe der
Wahrscheinlichkeitsrechnung.} Springer, Berlin.

\bibitem{KE} Kolmogorov, A. N. (1956). \textit{Foundations of the Theory of
Probability}. Chelsea Publ. Company, New York.

\bibitem{C9}
Lane, D. (2009). ‘Coloured Revolution’ as a Political Phenomenon. \textit{Journal of Communist Studies and Transition Politics}, 25(2), 113–135.

\bibitem{Lum} Lum, P.Y., Singh, G., Lehman, A., Ishkanov, T., Vejdemo-Johansson, M., Alagappan, M., Carlsson, J., Carlsson, G.
(2013). Extracting insights from the shape of complex data using topology. Scientific Reports  3, 1236

\bibitem{Majorana1} Majorana, E. (1942). Il valore delle leggi statistiche nella fisica e nelle scienze sociali. Scientia, 36(4), 58–66.
\bibitem{Majorana2} Majorana, E. (2006). The Value of Statistical Laws in Physics and Social Sciences. In L. Cifarelli (Ed.), Scientific Papers of Ettore Majorana (pp. 237-260). Cham: Springer

\bibitem{C10}
Mehdıyev, N. (2020). The Media and Social Networks as Factors in the ‘Colour Revolutions’. \textit{Technium Social Sciences Journal}, 11(1), 364–370.


\bibitem{Melkikh1} Melkikh, A. V. (2013). Biological complexity, quantum coherent states and the problem of efficient transmission of information inside a cell. BioSystems, 111(3), 190-198.
\bibitem{Melkikh2} Melkikh, A. V. (2024). Why does a cell function? New arguments in favor of quantum effects. BioSystems, 245, 105311.
\bibitem{Melkikh3} Melkikh, A. V. (2023). Thinking, holograms, and the quantum brain. Biosystems, 229, 104926.

\bibitem{Moreno1} Moreno, J. L., Organization of the Social Atom.
Sociometric Review, 1(1), 11-16 (1936). Reprinted in 
Sociometry, Experimental Method, and the Science of Society, Beacon House, New York, 1951, pp. 58-64. 
\bibitem{Moreno2} Moreno, J. L. The Social Atom and Death. Sociometry, 10(1), 80-84 (1947).
\bibitem{Moreno3} Moreno, J. L. Who Shall Survive? A New Approach to the Problem of Human Interrelations.
Washington, DC: Nervous and Mental Disease Pub. Co., 1934. Revised edition: Beacon House, New York, 1953. 
\bibitem{Moreno4} Moreno, J. L. Sociometry, Experimental Method, and the Science of Society: An Approach to a New Political Orientation.
Beacon House, New York, 1951.
\bibitem{Moreno5} Moreno, J. L. (ed.) The Sociometry Reader. Free Press (Beacon House imprint), New York, 1960. 
\bibitem{Moreno6} Moreno, J. L., \& Jennings, H. H. Statistics of Social Configurations. Sociometry 1, 342-374 (1938) 
\bibitem{Moreno7} Moreno, J. L., with discussion by W. A. White. Psychological Organization of Groups in the Community.
In 57th Yearbook on Mental Deficiency, pp. 3-25. Boston, MA: Association for Mental Deficiency, 1933. Revised edition available in Sociometry, Experimental Method… (1951) 


\bibitem{OK20} Ozawa, M., \& Khrennikov, A. (2020). Application of theory of
quantum instruments to psychology: Combination of question order effect with
response replicability effect. \textit{Entropy}, 22(1), 37.1--9436.

\bibitem{OK21} Ozawa M., \& Khrennikov A. (2021). Modeling combination of
question order effect, response replicability effect, and QQ-equality with
quantum instruments. \emph{Journal of Mathematical Psychology}, 100, 102491.

\bibitem{OK23} Ozawa M., \& Khrennikov A. (2023). Nondistributivity of human
logic and violation of response replicability effect in cognitive
psychology. \emph{Journal of Mathematical Psychology}, 112, 102739.

\bibitem{P} Penrose, R. (1989). \emph{The Emperor's New Mind}. Oxford
University Press: New-York.

\bibitem{Pothos1} Pothos, E.M.; Busemeyer, J.R. (2022). Quantum Cognition. 
\textit{Annual Review of Psychology}. 73, 749-778.

\bibitem{Pothos2} Epping, G. P., Fisher, E. L., Zeleznikow‐Johnston, A. M., Pothos, E. M., \& Tsuchiya, N. (2023). A quantum geometric framework for modeling color similarity judgments. Cognitive Science, 47(1), e13231.

\bibitem{PL1} Plotnitsky, A.  (2016).  {\it The principles of quantum theory, from Planck's quanta to the Higgs boson: The nature of quantum reality and the spirit of Copenhagen.}  Springer: Heidelberg-Berlin-New York. 
\bibitem{PL2} Plotnitsky, A. Spooky predictions at a distance: reality, complementarity and contextuality in quantum theory.
{\it Phil. Trans. R. Soc.} A  {\bf 2019}, {\it 377}, 20190089.
\bibitem{PL3} A. Plotnitsky, Reality without Realism. Matter, Thought, and Technology in Quantum Physics. Springer, Berlin-Heidelberg-New York, 2021.
\bibitem{PL4} Plotnitsky, A. (2023). The agency of observation not to be neglected’: complementarity, causality and the arrow of events in quantum and quantum-like theories. Phil. Trans. Royal Soc. A, 381(2256), 20220295.


\bibitem{Zeit} Plotnitsky, A.; Haven, E. (2023). \textit{The Quantum-Like
Revolution. A Festschrift for Andrei Khrennikov with a foreword by 2022
Nobel Laureate Anton Zeilinger.} Springer Nature, Berlin.

\bibitem{laser10} Rodin, L., Khrennikov, A. (2023). \textit{Social Laser:
Media and Social Mobilization}.
https://www.preprints.org/manuscript/202312.1570/v1

\bibitem{Savage} Savage L.J. (1954). \textit{The Foundations of Statistics}.
Wiley and Sons, New York.

\bibitem{SchT} Schr\"{o}dinger, E. (1989). \emph{Statistical thermodynamics}, Dover Publications.


\bibitem{shafir1992}
Shafir, E., \& Tversky, A. (1992).
Thinking through uncertainty: Nonconsequential reasoning and choice.
\textit{Cognitive Psychology}, 24(4), 449--474.

\bibitem{shafir1993}
Shafir, E., Simonson, I., \& Tversky, A. (1993).
Reason-based choice.
\textit{Cognition}, 49(1--2), 11--36.

\bibitem{shafir1993b}
Shafir, E. (1993).
Choosing versus rejecting: Why some options are both better and worse than others.
\textit{Memory \& Cognition}, 21(4), 546--556.

\bibitem{C11}
Silitski, V. (2009). What Are We Trying to Explain? \textit{Journal of Democracy}, 20(1), 86–89.


\bibitem{laser11} Tsarev, D., Trofimova, A., Alodjants, A., \& Khrennikov,
A. (2019). Phase transitions, collective emotions and decision-making
problem in heterogeneous social systems. \textit{Scientific reports}, 9(1),
18039. 

\bibitem{Tarde1} G. de Tarde, Les lois de l'imitation: etude sociologique. 1st ed. 1890; 2nd ed. 1895. Publ. Alcan, Paris
Reprints: Paris: Kime Editeur, 1993 (2nd ed., 428 pages) 
Les lois sociales: Esquisse d`une sociologie (Social Laws: An Outline of Sociology, 1898)
\bibitem{Tarde2} de Tarde, G. (1898). Les lois sociales: esquisse d`une sociologie. Publ.Alcan(Collection Biblioth`eque de philosophie contemporaine) Paris.

\bibitem{kahneman1981}
Tversky, A., \& Kahneman, D. (1981).
The framing of decisions and the psychology of choice.
\textit{Science}, 211(4481), 453--458.

\bibitem{tversky1992}
Tversky, A., \& Shafir, E. (1992).
The disjunction effect in choice under uncertainty.
\textit{Psychological Science}, 3(5), 305--309.


\bibitem{V1} Vitiello, G. (1995). Dissipation and memory capacity in the quantum brain model, \emph{Int. J. Mod. Phys.}, B9,  973.

\bibitem{V2} Vitiello, G. (2001).  \emph{ My double unveiled: The dissipative quantum model of brain}, Advances in Consciousness Research,  John Benjamins Publishing Company.

\bibitem{V3} Vallortigara, G.,  Vitiello, G. (2024). Brain asymmetry as minimization of free energy: a theoretical model. Royal Society Open Science, 11(7), 240465. 

\bibitem{VN} Von Neumann, J. {\it Mathematical Foundations of Quantum Mechanics}; Princeton University Press: Princeton, NJ, USA, 1955.

\bibitem{C12}
Way, L.A. (2008). The Real Causes of the Color Revolutions. \textit{Journal of Democracy}, 19(3), 55–69.

\bibitem{C13}
White, S. \& McAllister, I. (2014). Did Russia (Nearly) Have a Facebook Revolution in 2011? Social Media’s Challenge to Authoritarianism. \textit{Journal of Democracy}, 23(3), 63–70.

\bibitem{Winiarski1} Winiarski, L. (1904). Essais de Sociologie et d’Économie Sociales. Paris: P. Mouillot.
Early attempts to apply statistical methods to society, precursor of mathematical sociology.
\bibitem{Winiarski2} Winiarski, L. (1926). Traité d’Économie Sociale. Paris: Payot.

\bibitem{C14}
Wolchik, S.L. (2012). Putinism Under Siege: Can There Be a Color Revolution? \textit{Journal of Democracy}, 23(3), 63–70.

\end{thebibliography}
\end{document}